# Secure Source Coding with a Public Helper

Kittipong Kittichokechai*, Yeow-Khiang Chia†, Tobias J. Oechtering*, Mikael Skoglund*

and Tsachy Weissman‡

**Abstract**

We consider secure multi-terminal source coding problems in the presence of a public helper. Two main scenarios are studied: 1) source coding with a helper where the coded side information from the helper is eavesdropped by an external eavesdropper; 2) triangular source coding with a helper where the helper is considered as a public terminal. We are interested in how the helper can support the source transmission subject to a constraint on the amount of information leaked due to its public nature. We characterize the tradeoff between transmission rate, incurred distortion, and information leakage rate at the helper/eavesdropper in the form of a rate-distortion-leakage region for various classes of problems.

## I. INTRODUCTION

Nowadays the Internet is an essential part of our daily life. We rely on many online services which inevitably create huge amounts of information flow in the network. With this huge amount of information, the main tasks for network designers are to ensure that the data can be transmitted reliably and also securely across the network. The latter requirement is becoming increasingly acute, especially when sensitive information is involved. Let us imagine a network in which information flows from one node to another through a number of intermediate nodes. The system design generally makes use of these intermediate nodes to help the transmission. However, these nodes might be public devices or terminals which we cannot fully trust with access to significant amounts of our information. This scenario leads to a natural tradeoff between cooperation and secrecy in the system and motivates the study of secure communication and compression in the presence of a public helper.

In this work, we consider a secure lossy source coding problem involving a public helper under an information leakage rate constraint. The secure source coding problem is essentially a source coding problem with an additional secrecy constraint. For a given source sequence $X^n$, and some relevant information $W$ that is available to the eavesdropper or the public helper, the information leakage rate is defined as a normalized mutual information $\frac{1}{n}I(X^n;W)$. The solution to a secure lossy source coding problem is the optimal tradeoff between transmission rate, incurred distortion at the decoder, and information leakage rate at the eavesdropper in the form of a rate-distortion-leakage region. A summary of the problem settings and contributions of this work is given below (see also Table I).

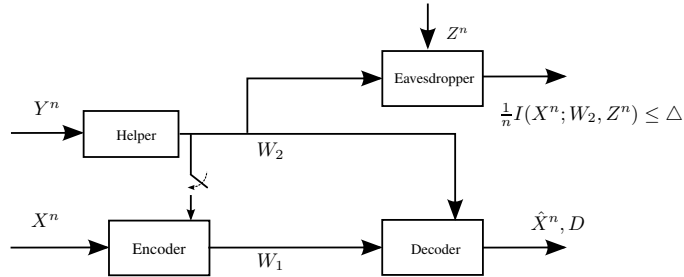

Fig. 1. Secure source coding with one-sided/two-sided public helper.

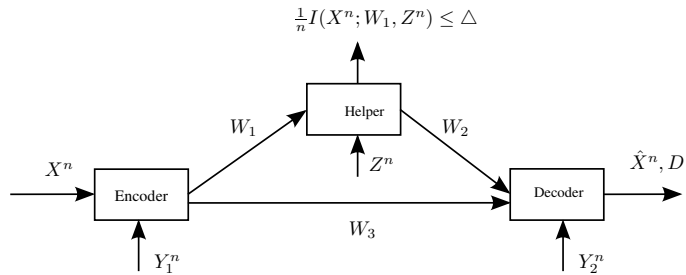

Fig. 2. Secure triangular/cascade source coding with public helper.

## A. Overview of Problem Setting and Contribution

*1) Secure Source Coding with a Public Helper:* First we consider secure lossy source coding with a public helper problems, as depicted in Fig. 1. The setting is motivated by a scenario where the helper can only provide the side information through a rate-limited communication link which is not secure due to its *public* nature, i.e., it can be eavesdropped by an external eavesdropper. In the "one-sided helper" setting, the helper communicates through a public link only to the decoder, while in the "two-sided helper" case, the helper *broadcasts* the same coded side information to both encoder and decoder. We provide an inner bound to the rate-distortion-leakage region for the one-sided helper case and show that it is tight under the logarithmic loss distortion measure, and for the Gaussian case with quadratic distortion and the Markov relation $Y - X - Z$. For the two-sided helper case, we solve the rate-distortion-leakage tradeoff under general distortion. We note that the one-sided/two-sided helper settings considered in Fig. 1 are essentially extensions of the one-helper problem [1–3] to the presence of an eavesdropper. Variation of the settings where the eavesdropper sees instead the link from an encoder to a decoder were studied in [4, 5].

*2) Secure Triangular/Cascade Source Coding with a Public Helper:* Next, we consider problems of triangular source coding with a public helper, as shown in Fig. 2. In contrast to the previous settings, where the focus is on leakage at an external eavesdropper, we address the problem of information leakage at a legitimate user. The setting is motivated by a scenario where the helper is a public terminal that forwards the information as the protocol requests from the encoder to the decoder. However, the helper might be curious and not ignore the data which may not be intended for him. The problem of characterizing the optimal rate-distortion-leakage tradeoff in general is difficult due to the ambiguity of the helper's strategy and the role of side information at the encoder. In this work, we characterize the rate-distortion-leakage regions for various special cases based on different side information patterns available to the encoder, the helper, and the decoder. Our contributions are summarized below.

- *Setting (A): We assume that $Y_1$ is constant, $Y_2 = Y$, and that $X - Y - Z$ forms a Markov chain. We solve the problem under the logarithmic loss distortion and for the Gaussian sources with quadratic distortion, and show that the forwarding scheme at the helper (setting $W_2 = W_1$) is optimal. Note that the Markov assumption $X - Y - Z$ in this setting can be relevant in scenarios where the decoder is a fusion center collecting all correlated side information.*
- *Setting (B): We assume that the side information $Y_1 = Y_2 = Y$, and that $X - Y - Z$ forms a Markov chain. We solve the problem under the logarithmic loss distortion and for the Gaussian source with quadratic distortion. We show that the availability of the side information at the encoder does not improve the rate-distortion tradeoff, and that the forwarding scheme at the helper is optimal. Interestingly, we note that although the availability of the side information at the encoder does not improve the rate-distortion tradeoff, this side information can be used for a secret key generation at the encoder and the decoder. In our coding scheme, the secret key is used to scramble part of the message sent to the helper, and thus decrease the information leakage.*
- *Setting (C): We assume that $(Y_1, Z)$ is constant and $Y_2 = Y$. It can be seen that the setting essentially reduces to the Wyner-Ziv like problem with an additional leakage constraint, and that the forwarding scheme at the helper is optimal. The Wyner-Ziv like coding achieves the whole rate-distortion-leakage region in this case.*
- *Setting (D): We assume that the side information at the helper is also available at the encoder, i.e., $Y_1 = Z$, and we let $Y_2 = Y$. In this case we assume that $X - Z - Y$ forms a Markov chain and solve the problem under general distortion. Due to $X - Z - Y$, we show that the decode-and-re encode type scheme at the helper is optimal. That is, it is meaningful to take into account $Z^n$ at the helper in relaying information to the decoder.*

We note that our settings are different from the conventional triangular/cascade source coding problem in that the decoding constraint at the helper is replaced by the secrecy constraint. Since all the cascade settings in our work can be seen as special cases of the triangular settings when the private link from the encoder to the decoder is removed (setting $W_3$ constant), we will provide the results and proofs for the triangular settings and state the cascade results as corollaries.

Apart from the triangular settings mentioned above, one might consider a slightly different setting where the encoder "broadcasts" the same source description to the helper and the decoder, as depicted in Fig. 3. Note that this is not a special case of previous triangular settings in Fig. 2 since it is more restrictive than simply setting the

|  | Source Coding with Public Helper | | Triangular/Cascade Source Coding with Public Helper | | | |
| --- | --- | --- | --- | --- | --- | --- |
|  | one-sided | two-sided | setting (A) | setting (B) | setting (C) | setting (D) |
| general distortion | ? | ✓ | ? | ? | ✓ | ✓(X-Z-Y) |
| logarithmic loss distortion | ✓ | ✓ | ✓(X-Y-Z) | ✓(X-Y-Z) | - | - |
| Gaussian w/ quadratic dist. | ✓(Y-X-Z) | ✓(Y-X-Z) | ✓(X-Y-Z) | ✓(X-Y-Z) | - | - |

TABLE I

SUMMARY OF OUR CONTRIBUTIONS; THE CHECK MARK ✓ DENOTES THE CASES THAT WE SOLVED IN THIS PAPER, WHILE THE QUESTION MARK ? DENOTES THE CASES THAT ARE LEFT OPEN; THE MARKOV ASSUMPTION ON THE SIDE INFORMATION IS STATED IF ANY.

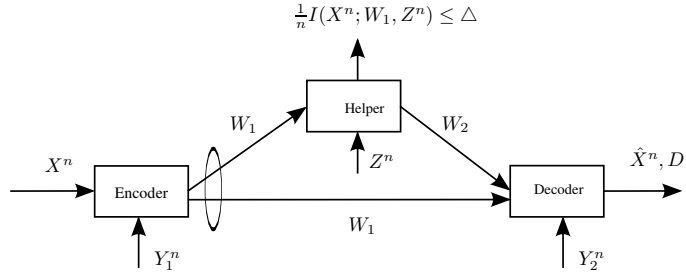

Fig. 3. Secure triangular source coding with public helper where the encoder "broadcasts" the same message.

rate $R_1 = R_3$. Depending on the side information pattern, we will see that, in some cases (setting (A)-(C)), the helper is not helpful in terms of providing more information to the decoder, i.e., $R_2 \geq 0$ is achievable.

### B. Related Work

Multi-terminal source coding problems have been studied extensively in various settings. The lossless distributed source coding problem was solved by Slepian and Wolf [6]. The setting has since been extended to the lossy case (cf., e.g., [7]), and remains open in general. It is also unsolved even when we require only one distortion constraint, i.e., one-helper problem [1], [2]. There exist only a few special cases which can be solved completely. This includes the Wyner-Ziv problem [8], the case when one source is decoded losslessly [9], and the special case of a Gaussian source with quadratic distortion [10], [11]. Recently, Courtade and Weissman [12] introduced a logarithmic loss distortion as a new and interesting distortion measure for lossy distributed source coding and solved the problem completely under this distortion measure. As many of our settings are solved under the logarithmic loss distortion, its definition and important properties, taken from [12], are briefly given at the end of this section. As for other related multi-terminal source coding problems, Yamamoto in [13] studied and established the rate-distortion regions for the cascade and triangular source coding problems without side information. Variations of the cascade and triangular source coding settings have been studied in recent years (see, e.g., [14, 15]).

Traditionally, most works concentrated only on problems of reliable communication or compression under a distortion constraint. Recently, security has become an important issue in system design and received substantial attention, i.e., when the goal is to design a communication system that is both reliable and secure. Physical layer security was introduced based on the fact that the signals available at the legitimate receiver and the eavesdropper usually possess different characteristics. It has gained significant attention due to its advantages in some applications, for example, those where complex cryptographic protocols cannot be implemented. Information theoretic study of physical layer security was pioneered by Wyner in [16], which introduces and solves the problem of coding for the Wiretap channel. Wyner has shown that perfect secure communication is possible when the channel from the sender to the legitimate receiver is "stronger" than the channel to the eavesdropper. Later generalizations include the case of a broadcast channel with confidential messages by Csiszár and Körner [17]. Several extensions to other multiuser channels including secure coding for multiple access channels, channels with states, etc. are considered in [18]. More recently, due to potential applications in areas such as privacy in sensor networks and databases (see, e.g., [19]), and privacy of distributed storage of genomic data (see, e.g., [20], [21]), an idea of physical layer security from the source coding perspective was also studied, i.e., source coding with side information subject to an additional secrecy constraint. As before, security/privacy of the source relies mainly on the different characteristics of signals (side information) which are available at the legitimate receiver and the eavesdropper. Secure lossless distributed source coding was studied by Prabhakaran and Ramchandran [22], Gündüz et al. [23], and Tandon et al. [4], and the lossy case was recently considered by Villard and Piantanida [5] and Ekrem and Ulukus [24]. The closely related work which characterized the tradeoff between amplifying information about one source and masking another was recently studied in [25]. Another line of work considers *explicit* secret key sharing in the system model that is based on the Shannon cipher system [26–31]. We note that we do not assume any explicit secret key sharing in this work. Nevertheless, in some scenarios, we may be able to exploit some common randomness for secrecy by *implicitly* generating a secret key using the common side information (Section IV, setting (B)).

Below we give a brief definition and important properties of the logarithmic loss distortion. Logarithmic loss has the interesting property that, when used as a distortion measure in the Wyner-Ziv (like) problem [8], the side information at the encoder does not improve the rate-distortion region. This property is essential in establishing a couple of complete results in this paper by using the achievable schemes which neglect the side information at the encoder.

*Logarithmic Loss Distortion Measure* [12]: We let the reconstruction alphabet $\hat{\mathcal{X}}$ be the set of probability distribution over the source alphabet $\mathcal{X}$, i.e., $\hat{\mathcal{X}} := \{p : p \text{ is a pmf on } \mathcal{X}\}$. For a sequence $\hat{X}^n \in \hat{\mathcal{X}}^n$, we denote $\hat{X}_i$, $i = 1, \ldots, n$, the $i^{th}$ element of $\hat{X}^n$. Then $\hat{X}_i, i = 1, \ldots, n$ is a probability distribution on $\mathcal{X}$, i.e., $\hat{X}_i : \mathcal{X} \to [0, 1]$, and $\hat{X}_i(x)$ is a probability distribution on $\mathcal{X}$ evaluated for the outcome $x \in \mathcal{X}$. In other words, the decoder generates the "soft" estimates of the source sequence.

*Definition 1 (logarithmic loss):* The logarithmic loss distortion measure is defined as $d(x, \hat{x}) = \log(\frac{1}{\hat{x}(x)}) = D_{KL}(\mathbf{1}_{\{x\}} || \hat{x})$, where $\mathbf{1}_{\{x\}} : \mathcal{X} \to \{0, 1\}$ is an indicator function such that, for $a \in \mathcal{X}$, $\mathbf{1}_{\{x\}}(a) = 1$ if $a = x$, and $\mathbf{1}_{\{x\}}(a) = 0$ otherwise. That is, $d(x, \hat{x})$ is the Kullback-Leibler divergence between the empirical distribution of the event $X = x$ and the estimate $\hat{x}$. Using this definition for symbol-wise distortion, it is standard to define the distortion between sequences as $d^{(n)}(x^n, \hat{x}^n) = \frac{1}{n}\sum_{i=1}^{n} d(x_i, \hat{x}_i)$.

In the following we present a couple of lemmas which appear in [12] and are essential in proving our results under the logarithmic loss distortion. Lemma 1 is used in the achievability proof (inner bound argument), while Lemma 2 is used for upper bounding the conditional entropy in the converse proof (outer bound argument). Both follow quite directly from the definition of logarithmic loss, but we include their proofs [12] in the appendix for completeness.

*Lemma 1 (inner bound argument):* Let $U$ be the argument of the reconstruction function $g(\cdot)$, then under the logarithmic loss distortion measure, we get $E[d(X, g(U))] = H(X|U)$.

*Lemma 2 (outer bound argument):* Let $Z = (W_1, W_2)$ be the argument of the reconstruction function $g^{(n)}(\cdot)$, then under the logarithmic loss distortion measure, we get $E[d^{(n)}(X^n, g^{(n)}(Z))] \geq \frac{1}{n} H(X^n|Z)$.

## C. Organization

The remaining parts of the paper are organized as follows. In Section II we provide definitions and detailed problem formulations for secure source coding with one-sided/two-sided public helper, and secure triangular source coding with a public helper. Section III provides the rate-distortion-leakage tradeoff for secure source coding with one-sided/two-sided public helper, as depicted in Fig. 1. A Gaussian example for the one-sided helper setting is also given. In Section IV we characterize the rate-distortion-leakage regions for several special cases of secure triangular/cascade source coding with a public helper (settings (A)-(D) in Fig. 2, and the corresponding "broadcast" setting in Fig. 3).

*Notation*: We denote the discrete random variables, their corresponding realizations or deterministic values, and their alphabets by the upper case, lower case, and calligraphic letters, respectively. The term $X_m^n$ denotes the sequence $\{X_m, \ldots, X_n\}$ when $m \leq n$, and the empty set otherwise. Also, we use the shorthand notation $X^n$ for $X_1^n$. The term $X^{n \setminus i}$ denotes the set $\{X_1, \ldots, X_{i-1}, X_{i+1}, \ldots, X_n\}$. Cardinality of the set $\mathcal{X}$ is denoted by $|\mathcal{X}|$. Notation $[1 : 2^{nI(X;Y)}]$ denotes the set $\{1, 2, \ldots, 2^{nI(X;Y)}\}$. For $a \in \mathbb{R}$, $[a]^+$ is defined as $\max\{0, a\}$. Finally, we use $X - Y - Z$ to denote that $(X, Y, Z)$ forms a Markov chain, that is, their joint PMF factorizes as $P_{X,Y,Z}(x, y, z) = P_{X,Y}(x, y) P_{Z|Y}(z|y)$ or $P_{X,Y,Z}(x, y, z) = P_{X|Y}(x|y) P_{Y,Z}(y, z)$.

## II. PROBLEM SETTING

First, we consider source coding with a helper where the helper's link is public and can therefore be eavesdropped by an external eavesdropper, as depicted in Fig. 1. We state the detailed problem formulation below and provide the characterization of rate-distortion-leakage rate tradeoff in the form of a rate-distortion-leakage region in Section III.

### A. Secure Source Coding with One-sided Public Helper

Let us consider the setting in Fig. 1 when the switch is open. Source, side information, and reconstruction alphabets, $\mathcal{X}, \mathcal{Y}, \mathcal{Z}, \hat{\mathcal{X}}$ are assumed to be finite. Let $(X^n, Y^n, Z^n)$ be the n-length sequences which are i.i.d. according to $P_{X,Y,Z}$. Given a source sequence $X^n$, an encoder generates a source description $W_1 \in \mathcal{W}_1^{(n)}$ and sends it over the noise-free, rate-limited link to a decoder. Meanwhile, a helper who observes the side information $Y^n$ generates coded side information $W_2 \in \mathcal{W}_2^{(n)}$ and sends it to the decoder over another noise-free, rate-limited link. Given the source description and the coded side information, the decoder reconstructs the source sequence as $\hat{X}^n$ subject to a distortion constraint. We note that the eavesdropper also receives the coded side information and its own side information $Z^n$.

*Definition 2:* A $(|\mathcal{W}_1^{(n)}|, |\mathcal{W}_2^{(n)}|, n)$-code for secure source coding with one-sided public helper consists of:

- A stochastic encoder $F_1^{(n)}$ which takes $X^n$ as an input and generates $W_1 \in \mathcal{W}_1^{(n)}$ according to a conditional pmf $p(w_1|x^n)$,
- A stochastic helper $F_2^{(n)}$ which takes $Y^n$ as an input and generates $W_2 \in \mathcal{W}_2^{(n)}$ according to $p(w_2|y^n)$, and
- A decoder $g^{(n)} : \mathcal{W}_1^{(n)} \times \mathcal{W}_2^{(n)} \to \hat{\mathcal{X}}^n$,

where $\mathcal{W}_1^{(n)}$ and $\mathcal{W}_2^{(n)}$ are finite sets.

Let $d : \mathcal{X} \times \hat{\mathcal{X}} \to [0, \infty)$ be the single-letter distortion measure. The average distortion between the source sequence and its reconstruction at the decoder is defined as

$$E\left[d^{(n)}(X^n, \hat{X}^n)\right] \triangleq \frac{1}{n} E\left[\sum_{i=1}^{n} d(X_i, \hat{X}_i)\right],$$

where $d^{(n)}(\cdot)$ is the distortion function.

The information leakage rate at the eavesdropper who has access to $W_2$ and $Z^n$ is measured by the normalized mutual information $\frac{1}{n} I(X^n; W_2, Z^n)$.

*Definition 3:* The rate-distortion-leakage tuple $(R_1, R_2, D, \triangle) \in \mathbb{R}_+^4$ is said to be *achievable* if for any $\delta > 0$ and all sufficiently large $n$ there exists a $(|\mathcal{W}_1^{(n)}|, |\mathcal{W}_2^{(n)}|, n)$ code such that

$$\frac{1}{n} \log |\mathcal{W}_i^{(n)}| \leq R_i + \delta, \ i = 1, 2,$$

$$E[d^{(n)}(X^n, g^{(n)}(W_1, W_2))] \leq D + \delta,$$

$$\text{and} \quad \frac{1}{n} I(X^n; W_2, Z^n) \leq \triangle + \delta.$$

The *rate-distortion-leakage region* $\mathcal{R}_{\text{one-sided}}$ is the set of all achievable tuples.

### B. Secure Source Coding with Two-sided Public Helper

Let us consider the setting in Fig. 1 when the switch is closed. Since the problem setting is similar to that of the one-sided helper case, details are omitted. The main difference is that the coded side information $W_2 \in \mathcal{W}_2^{(n)}$ is given to both the encoder and the decoder. Then, based on $X^n$ and $W_2$, the encoder generates the source description $W_1 \in \mathcal{W}_1^{(n)}$. That is, the encoding function becomes $F_1^{(n)}$ that takes $(X^n, W_2)$ as input and generates $W_1$ according to $p(w_1|x^n, w_2)$.

Next, we consider triangular/cascade source coding settings in which the helper plays a role in relaying information from an encoder to a decoder subject to the leakage constraint, as depicted in Fig. 2. Clearly, there exists a tradeoff between amount of information leakage to the helper and the helper's ability to support the source transmission. We state general problem formulation below and characterize the rate-distortion-leakage rate tradeoff in the form of a rate-distortion-leakage region for settings (A)-(D) in Section IV.

*C. Secure Triangular/Cascade Source Coding with a Public Helper*

Let us consider the setting in Fig. 2. Source and side information sequences $(X^n, Y_1^n, Y_2^n, Z^n)$ are assumed to be i.i.d. according to $P_{X,Y_1,Y_2,Z}$. Given the sequences $(X^n, Y_1^n)$, an encoder generates a description $W_1 \in \mathcal{W}_1^{(n)}$ and sends it to the helper over a noise-free, rate-limited link. The encoder also generates a description $W_3 \in \mathcal{W}_3^{(n)}$ based on $(X^n, Y_1^n)$ and sends it to the decoder over another noise-free, rate-limited link. Based upon the description $W_1$ and the side information $Z^n$, the helper generates a new description $W_2 \in \mathcal{W}_2^{(n)}$ and sends it to the decoder. Given $W_2, W_3$, and its own side information $Y_2^n$, the decoder reconstructs the source sequence as $\hat{X}^n$.

*Definition 4:* A $(|\mathcal{W}_1^{(n)}|, |\mathcal{W}_2^{(n)}|, |\mathcal{W}_3^{(n)}|, n)$-code for secure triangular source coding with a public helper consists of:

- A stochastic encoder $F_1^{(n)}$ which takes $(X^n, Y_1^n)$ as input and generates $W_1 \in \mathcal{W}_1^{(n)}$ according to a conditional pmf $p(w_1|x^n, y_1^n)$,
- A stochastic helper $F_2^{(n)}$ which takes $(W_1, Z_1^n)$ as input and generates $W_2 \in \mathcal{W}_2^{(n)}$ according to $p(w_2|w_1, z^n)$,
- A stochastic encoder $F_3^{(n)}$ which takes $(X^n, Y_1^n)$ as input and generates $W_3 \in \mathcal{W}_3^{(n)}$ according to $p(w_3|x^n, y_1^n)$, and
- A decoder $g^{(n)} : \mathcal{W}_2^{(n)} \times \mathcal{W}_3^{(n)} \times \mathcal{Y}_2^{(n)} \to \hat{\mathcal{X}}^n$,

where $\mathcal{W}_1^{(n)}, \mathcal{W}_2^{(n)}$, and $\mathcal{W}_3^{(n)}$ are finite sets.

The information leakage rate at the helper who has access to $W_1$ and $Z^n$ is measured by $\frac{1}{n} I(X^n; W_1, Z^n)$.

*Definition 5:* The rate-distortion-leakage tuple $(R_1, R_2, R_3, D, \triangle) \in \mathbb{R}_+^5$ is said to be *achievable* if for any $\delta > 0$ and all sufficiently large $n$ there exists a $(|\mathcal{W}_1^{(n)}|, |\mathcal{W}_2^{(n)}|, |\mathcal{W}_3^{(n)}|, n)$ code such that

$$\frac{1}{n} \log |\mathcal{W}_i^{(n)}| \leq R_i + \delta, \ i = 1, 2, 3,$$

$$E[(X^n, g^{(n)}(W_2, W_3, Y_2^n))] \leq D + \delta,$$

$$\text{and } \frac{1}{n} I(X^n; W_1, Z^n) \leq \triangle + \delta.$$

The *rate-distortion-leakage* region is defined as the set of all achievable tuples.

The problem formulations for the cascade settings follow straightforwardly as special cases of the triangular settings.

III. SECURE SOURCE CODING WITH ONE-SIDED/TWO-SIDED PUBLIC HELPER

In this section, we provide the characterizations of the rate-distortion-leakage regions for secure source coding with one-sided/two-sided public helper. For the one-sided public helper case, we first provide the inner bound to the rate-distortion-leakage region. Then we show that this inner bound is tight for some special cases including the cases under the logarithmic loss distortion measure, and the Gaussian case with quadratic distortion under $Y - X - Z$ Markov assumption. Next, we consider the two-sided helper case and characterize the rate-distortion-leakage region under a general distortion function.

9## A. One-sided Public Helper

*Theorem 1 (Inner Bound):* The rate-distortion-leakage region $\mathcal{R}_{\text{one-sided}}$ *contains* the convex closure of the set of all tuples $(R_1, R_2, D, \triangle) \in \mathbb{R}_+^4$ for which there exist random variables $U \in \mathcal{U}$, and $V \in \mathcal{V}$ such that $U - Y - (X, Z, V)$ and $V - X - (U, Y, Z)$ form Markov chains and a function $g : \mathcal{U} \times \mathcal{V} \to \mathcal{X}$ that satisfy

$$R_2 \geq I(Y;U),$$
$$R_1 \geq I(X;V|U),$$
$$D \geq E[d(X, g(U,V))],$$
$$\triangle \geq I(X;U,Z).$$

The cardinalities of the alphabets of the auxiliary random variables can be upperbounded as $|\mathcal{U}| \leq |\mathcal{Y}| + 4$, $|\mathcal{V}| \leq |\mathcal{X}| + 1$.

*Remark 1:* Our achievable scheme is identical to that of the original one-helper problem. However, the resulting rate-distortion tradeoff is different since the set of optimizing input distributions may change due to an additional leakage constraint. We note also that the problem of characterizing the complete rate-distortion-leakage region under general distortion remains open. This is to be expected in view of the fact that the one-helper problem (without the leakage rate constraint) is still open.

*Proof of Theorem 1*: The proof is based on the random coding argument. The achievable scheme follows the standard rate-distortion and Wyner-Ziv like coding scheme. That is, for fixed $P_{U|Y}, P_{V|X}$, and $g(\cdot)$, randomly generate $2^{n(I(Y;U)+\epsilon)}$ sequences $u^n(w_2) \sim \prod_{i=1}^n P_U(u_i(w_2)), w_2 \in [1 : 2^{n(I(Y;U)+\epsilon)}]$. Also, randomly generate $2^{n(I(X;V)+\epsilon)}$ sequences $v^n(\tilde{w}) \sim \prod_{i=1}^n P_V(v_i(\tilde{w})), \tilde{w} \in [1 : 2^{n(I(X;V)+\epsilon)}]$, and distribute them uniformly into $2^{n(I(X;V|U)+2\epsilon)}$ bins $b_v(w_1), w_1 \in [1 : 2^{n(I(X;V|U)+2\epsilon)}]$. For encoding, the helper looks for $u^n$ that is jointly typical with $y^n$. If there is more than one, it selects one of them uniformly at random. If there is no such $u^n$, it selects one out of $2^{n(I(Y;U)+\epsilon)}$ uniformly at random. Then it transmits the corresponding index $w_2$ to the decoder. With high probability, there exists such $u^n$ since there are $2^{n(I(Y;U)+\epsilon)}$ codewords generated. The encoder looks for $v^n$ that is jointly typical with $x^n$. If there is more than one, it selects one of them uniformly at random. If there is no such $v^n$, it selects one out of $2^{n(I(X;V)+\epsilon)}$ uniformly at random. It then transmits the corresponding bin index $w_1$ to the decoder. With high probability, there exists such $v^n$ since there are $2^{n(I(X;V)+\epsilon)}$ codewords generated. Upon receiving $(w_1, w_2)$, the decoder looks for the unique $v^n$ such that it is jointly typical with $u^n$. With high probability, it will find the unique and correct one since there are $2^{n(I(U;V)-\epsilon)}$ codewords in each bin $b_v(w_1)$. Then $\hat{x}^n$ is put out as a source reconstruction, where $\hat{x}_i = g(u_i, v_i)$. Since $(x^n, u^n, v^n)$ are jointly typical, we can show that $D \geq E[d(X, g(U,V))]$ is achievable.

As for the analysis of leakage rate, we consider the normalized mutual information averaged over all codebooks $\mathcal{C}_n$,

$$I(X^n; W_2, Z^n | \mathcal{C}_n)$$
$$= H(X^n | \mathcal{C}_n) - H(X^n, W_2, Z^n | \mathcal{C}_n) + H(W_2, Z^n | \mathcal{C}_n)$$
$$= H(X^n | \mathcal{C}_n) - H(X^n, Z^n | \mathcal{C}_n) - H(W_2 | X^n, Z^n, \mathcal{C}_n) + H(W_2 | \mathcal{C}_n) + H(Z^n | W_2, \mathcal{C}_n)$$
$$\leq H(X^n | \mathcal{C}_n) - H(X^n, Z^n | \mathcal{C}_n) - I(W_2; Y^n | X^n, Z^n, \mathcal{C}_n) + H(W_2 | \mathcal{C}_n) + H(Z^n | W_2, \mathcal{C}_n)$$
$$\stackrel{(a)}{=} H(X^n) - H(X^n, Y^n, Z^n) + H(Y^n | W_2, X^n, Z^n, \mathcal{C}_n) + H(W_2 | \mathcal{C}_n) + H(Z^n | W_2, \mathcal{C}_n)$$
$$\stackrel{(b)}{\leq} n[H(X) - H(X, Y, Z) + H(Y | U, X, Z) + \delta_\epsilon + I(Y; U) + \epsilon + H(Z | U) + \delta_\epsilon]$$
$$\stackrel{(c)}{=} n[H(X) - H(X, Z | Y, U) - I(Y; X, Z | U) + H(Z | U) + \delta'_\epsilon]$$
$$= n[I(X; U, Z) + \delta'_\epsilon]$$

where $(a)$ follows from the facts that $(X^n, Y^n, Z^n)$ are independent of the codebook, $(b)$ follows from the i.i.d. property of $(X^n, Y^n, Z^n)$, from the codebook generation that we have $W_2 \in [1 : 2^{n(I(Y;U)+\epsilon)}]$, and from Lemma 3 and 4 below in which we bound the terms $H(Y^n | W_2, X^n, Z^n, \mathcal{C}_n)$ and $H(Z^n | W_2, \mathcal{C}_n)$, and that $\Pr((Y^n, U^n(W_2), X^n, Z^n) \in \mathcal{T}_\epsilon^{(n)}) \to 1$ as $n \to \infty$ from the codebook generation and encoding process, $(c)$ follows from the Markov chain $U - Y - (X, Z)$, and that $\delta'_\epsilon := \epsilon + 2\delta_\epsilon \to 0$ as $\epsilon \to 0$.

*Lemma 3:* Let $W_2$ be the corresponding index of codeword $U^n$. If $\Pr((Y^n, U^n(W_2), X^n, Z^n) \in \mathcal{T}_\epsilon^{(n)}) \to 1$ as $n \to \infty$, we have that $\frac{1}{n} H(Y^n | W_2, X^n, Z^n, \mathcal{C}_n) \leq H(Y | U, X, Z) + \delta_\epsilon$

*Proof:* The proof is given in Appendix C. ∎

*Lemma 4:* Let $W_2$ be the corresponding index of codeword $U^n$. If $\Pr((U^n(W_2), Z^n) \in \mathcal{T}_\epsilon^{(n)}) \to 1$ as $n \to \infty$, we have that $\frac{1}{n} H(Z^n | W_2, \mathcal{C}_n) \leq H(Z | U) + \delta_\epsilon$.

*Proof:* The proof follows similarly as that of Lemma 3. ∎

For the bounds on the cardinalities of the sets $\mathcal{U}$ and $\mathcal{V}$, it can be shown by using the support lemma [32] that it suffices that $\mathcal{U}$ should have $|\mathcal{Y}| - 1$ elements to preserve $P_Y$, plus five more for $H(Y|U), H(X|U), H(X|U, V), H(X|U, Z)$, and the distortion constraint. And similarly, it suffices that $\mathcal{V}$ should have at most $|\mathcal{X}| + 1$ elements to preserve $P_X$, $H(X, U, V)$, and the distortion constraint. This finally concludes the proof. ∎

Next, we show that the inner bound provided in Theorem 1 is tight for some special cases, namely, the setting of the logarithmic loss distortion measure, and the Gaussian setting under quadratic distortion and the Markov assumption $Y - X - Z$.

*1) Logarithmic Loss Distortion:*

*Theorem 2 (Logarithmic Loss):* The rate-distortion-leakage region under logarithmic loss distortion $\mathcal{R}_\text{one-sided, logloss}$ is the set of all tuples $(R_1, R_2, D, \triangle) \in \mathbb{R}_+^4$ for which there exists a random variable $U \in \mathcal{U}$ such that $U - Y - (X, Z)$

forms a Markov chain and

$$R_2 \geq I(Y;U),$$
$$R_1 \geq [H(X|U) - D]^+,$$
$$\triangle \geq I(X;U,Z).$$

The cardinality of the alphabet of the auxiliary random variable can be upperbounded as $|\mathcal{U}| \leq |\mathcal{Y}| + 2$.

*Remark 2:* Interestingly, Theorem 2 shows that the achievable scheme for the original one-helper problem (the one used in Theorem 1) is also optimal in the presence of an eavesdropper. Due to the property of logarithmic loss distortion which allows the use of Lemmas 1 and 2 in proving achievability and converse, an additional auxiliary random variable $V$ and its associated Markov chain are not needed in characterizing the rate-distortion-leakage region. This essentially allows us to overcome the common issue faced in establishing a complete result for the lossy multi-terminal source coding problem in general.

*Proof of Theorem 2:*

*Sketch of Achievability*: The achievable proof follows the proof of the inner bound in Theorem 1. That is, the scheme consists of the rate-distortion code for lossy transmission of $y^n$ via the codeword $u^n$ at rate $I(Y;U) + \epsilon$, and the Wyner-Ziv code at rate $I(X;V|U) + 2\epsilon$ for lossy transmission of $x^n$ with $u^n$ as side information at the decoder. We can show that the distortion $D$ and the leakage $\triangle$, satisfying $D \geq E[d(X, g(U,V))], \triangle \geq I(X;U,Z)$, are achievable. Due to the property of logarithmic loss distortion function (Lemma 1), we have $E[d(X, g(U,V))] = H(X|U,V)$. If $H(X|U) < D$, the encoder does not need to send anything, i.e., setting $V$ constant. If $H(X|U) > D$, we define $V = X$ with probability $p = 1 - \frac{D}{H(X|U)}$ and constant otherwise. Then we get $H(X|U,V) = D$ and $I(X;V|U) = H(X|U) - D$. Therefore, we obtain the desired achievable rate-distortion-leakage expressions. The converse proof uses the fact that for logarithmic loss distortion function $E[d(X^n, g(W_1, W_2))] \geq \frac{1}{n}H(X^n|W_1, W_2)$ (Lemma 2), and it is given in Appendix D. ∎

*Remark 3:* We note that the proof of the inner bound in Theorem 1 holds only for bounded distortion measures. However, the logarithmic loss distortion measure is not bounded. To address this issue, we refer to an earlier version of [12] and also [33, Remark 3.4] where the proof of achievability can be extended to logarithmic loss distortion by perturbing the reconstruction probability distribution. That is, we assign a small positive value to the reconstruction probability distribution that takes value zero. By this perturbation, the maximum distortion incurred can be upperbounded, and the proof of the inner bound in Theorem 1 can then be applied with this perturbed reconstruction function.

*2) Gaussian Setting under Quadratic Distortion, and the Markov Relation $Y - X - Z$:* In this part, we evaluate the rate-distortion regions when $(X^n, Y^n, Z^n)$ are jointly Gaussian and the distortion function is quadratic. Let the sequences $(X^n, Y^n, Z^n)$ be i.i.d. according to $P_{X,Y,Z}$. We will assume that $Y \sim \mathcal{N}(0, \sigma_Y^2)$, $X = Y + N_1, N_1 \sim \mathcal{N}(0, \sigma_{N_1}^2)$ independent of $Y$, and $Z = X + N_2, N_2 \sim \mathcal{N}(0, \sigma_{N_2}^2)$ independent of $(X, Y, N_1)$, where $\sigma_Y^2, \sigma_{N_1}^2, \sigma_{N_2}^2 > 0$. Note that this satisfies the Markov assumption $Y - X - Z$. While our main results in previous cases were proven only for discrete memoryless sources, the extension to the quadratic Gaussian case is standard and it follows, for example, [34] and [35].

*Theorem 3 (Gaussian, $Y - X - Z$):* The rate-distortion-leakage region for a Gaussian source with quadratic distortion under the Markov assumption $Y - X - Z$, $\mathcal{R}_{\text{one-sided,Gaussian}}$, is the set of all tuples $(R_1, R_2, D, \triangle) \in \mathbb{R}_+^4$ that satisfy

$$R_2 \geq \frac{1}{2} \log (1/\alpha),$$
$$R_1 \geq \frac{1}{2} \log \Big(\frac{\alpha \sigma_Y^2 + \sigma_{N_1}^2}{D}\Big),$$
$$\triangle \geq \frac{1}{2} \log \Big(\frac{(\sigma_Y^2 + \sigma_{N_1}^2)(\alpha \sigma_Y^2 + \sigma_{N_1}^2 + \sigma_{N_2}^2)}{\sigma_{N_2}^2 (\alpha \sigma_Y^2 + \sigma_{N_1}^2)}\Big),$$

for some $\alpha \in (0, 1)$.

*Proof:* The proof is given in Appendix E. ∎

*Corollary 1:* The minimum achievable distortion for given rates and leakage rate $R_1, R_2, \triangle$ under the Markov assumption $Y - X - Z$ is given by

$$D_{\min}(R_1, R_2, \triangle) = \max\{2^{-2R_1}(2^{-2R_2} \sigma_Y^2 + \sigma_{N_1}^2), 2^{-2R_1}(\alpha^* \sigma_Y^2 + \sigma_{N_1}^2)\}, \quad (1)$$

where $0 \leq \alpha^* = \frac{2^{-2\triangle}(\sigma_Y^2 + \sigma_{N_1}^2)(\sigma_{N_1}^2 + \sigma_{N_2}^2) - \sigma_{N_1}^2 \sigma_{N_2}^2}{\sigma_Y^2 \sigma_{N_2}^2 - 2^{-2\triangle} \sigma_Y^2 (\sigma_Y^2 + \sigma_{N_1}^2)} = \frac{\sigma_{N_2}^2}{\left(\frac{2^{2\triangle} \sigma_{N_2}^2}{(\sigma_Y^2 + \sigma_{N_1}^2)} - 1\right) \sigma_Y^2} - \frac{\sigma_{N_1}^2}{\sigma_Y^2} \leq 1$, and $1/2 \log(1 + \frac{\sigma_Y^2 + \sigma_{N_1}^2}{\sigma_{N_2}^2}) \leq \triangle \leq 1/2 \log(\frac{(\sigma_Y^2 + \sigma_{N_1}^2)(\sigma_{N_1}^2 + \sigma_{N_2}^2)}{\sigma_{N_1}^2 \sigma_{N_2}^2})$.

*Proof:* The proof follows from the result in Theorem 3 where we use the fact that $1/2 \log(1 + \frac{\sigma_Y^2 + \sigma_{N_1}^2}{\sigma_{N_2}^2}) = I(X; Z) \leq \triangle \leq I(X; Y, Z) = 1/2 \log(\frac{(\sigma_Y^2 + \sigma_{N_1}^2)(\sigma_{N_1}^2 + \sigma_{N_2}^2)}{\sigma_{N_1}^2 \sigma_{N_2}^2})$, and solve for $D$. ∎

*Example 1:* We evaluate the minimum achievable distortion for given rates and leakage rate in Corollary 1. For fixed $\sigma_Y^2 = 0.5, \sigma_{N_1}^2 = \sigma_{N_2}^2 = 0.2$, we plot $D_{\min}$ as a function of $\triangle$ for given $R_1$ and $R_2$ in Fig 4.

We can see that, in general, for given $R_1$ and $R_2$, $D_{\min}$ is decreasing when $\triangle$ becomes larger. This is because the helper is able to transmit more information to the decoder without violating the leakage constraint. However, there exists a $\triangle^*$ such that for any $\triangle > \triangle^*$ we cannot improve $D_{\min}$ further by increasing $\triangle$ since it is limited by the rate $R_2$. This saturation effect can be seen from the expression of $D_{\min}$ as a max function in (1). That is, for given $R_1, R_2$, when $\triangle$ is sufficiently large, we get $D_{\min} = 2^{-2R_1}(2^{-2R_2} \sigma_Y^2 + \sigma_{N_1}^2)$ which is constant. In fact, we can determine the value of $\triangle^*$ from (1) by solving for $\triangle$ in the equation $2^{-2R_2} = \alpha^*$. We note that $\triangle^*$ depends only on $R_2$ as seen also from Fig. 4 that when $R_2 = 1.25$, we get the same $\triangle^*$ for different $R_1$, e.g., $R_1 = 1$ or 1.25. We note that $D_{\min}$ still depends on $R_1$, i.e., it is saturated at a lower level for larger $R_1$. To this end, we conclude that at high $\triangle$ region, $R_2$ is a limiting factor of $D_{\min}$.

On the other hand, when $\triangle$ is "small," the decreasing region is active, i.e., $D_{\min} = 2^{-2R_1}(\alpha^* \sigma_Y^2 + \sigma_{N_1}^2)$, and $D_{\min}$ depends only on $R_1$ and $\triangle$ (not on $R_2$). That is, in the "small" $\triangle$ region, $D_{\min}$ is limited by $\triangle$ so that we cannot improve $D_{\min}$ by increasing $R_2$ further. This can be seen from the plots that, for a given $R_1$, three distortion-leakage curves with different $R_2$ coincide in the small $\triangle$ region.

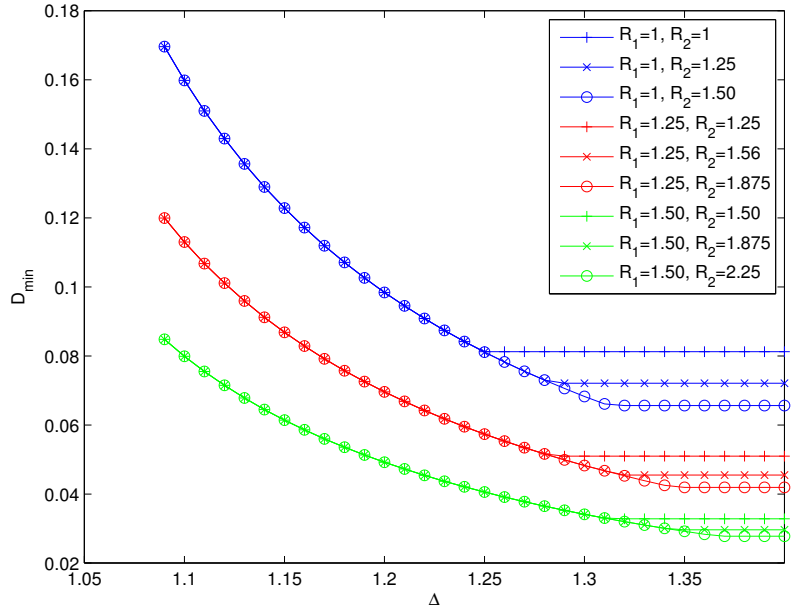

Fig. 4. Gaussian example: Minimum achievable distortion as a function of leakage rate for given rates $R_1, R_2$.

## B. Two-sided Public Helper

In this section, we characterize the rate-distortion-leakage region under general distortion function for the secure source coding with two-sided public helper problem.

*Theorem 4:* The rate-distortion-leakage region $\mathcal{R}_{\text{two-sided}}$ is the set of all tuples $(R_1, R_2, D, \triangle) \in \mathbb{R}_+^4$ for which there exist random variables $U \in \mathcal{U}$ and $\hat{X} \in \hat{\mathcal{X}}$ such that $U - Y - (X, Z)$ and $\hat{X} - (U, X) - (Y, Z)$ form Markov chains and

$$R_2 \geq I(Y; U),$$
$$R_1 \geq I(X; \hat{X}|U),$$
$$D \geq E[d(X, \hat{X})],$$
$$\triangle \geq I(X; U, Z). \qquad (2)$$

The cardinality of the alphabet of the auxiliary random variable can be upperbounded as $|\mathcal{U}| \leq |\mathcal{Y}| + 3$.

*Proof:* The achievable scheme consists of the rate-distortion code for lossy transmission of $y^n$ via $u^n$ at rate $I(Y; U) + \epsilon$. Since $w_2$ is given to both the encoder and the decoder, source coding with side information known at both encoder and decoder at rate $I(X; \hat{X}|U) + 2\epsilon$ is used for lossy transmission of $x^n$ with $u^n$ as side information. The achievable leakage rate proof and the converse proof follow similarly as that of one-sided helper case, and are therefore omitted. Similarly as in [3], we note that by following the converse proof of one-sided helper case, we in fact proved the outer bound which has the same rate, distortion, and leakage rate constraints as in (2), but with the joint distribution satisfying $U - Y - (X, Z)$ and $\hat{X} - (U, X, Y) - Z$. Clearly this outer bound includes the achievable region due to the larger set of distributions. To show that the outer bound is also included in the achievable region, we let $(R_1, R_2, D, \triangle)$ be in the outer bound with the joint distribution of the form

$$\bar{p}(x, y, z, u, \hat{x}) = p(x, y, z)p(u|y)\bar{p}(\hat{x}|u, x, y). \tag{3}$$

Then we show that there exists a distribution of the form satisfying the Markov conditions in the achievable region such that the constraints on $(R_1, R_2, D, \triangle)$ in (2) hold.

Let

$$p(x, y, z, u, \hat{x}) = p(x, y, z)p(u|y)\bar{p}(\hat{x}|u, x), \tag{4}$$

where $\bar{p}(\hat{x}|u, x)$ is induced by $\bar{p}(x, y, z, u, \hat{x})$. We now show that the terms $I(Y; U)$, $I(X; \hat{X}|U)$, $E[d(X, \hat{X})]$, and $I(X; U, Z)$ are the same whether we evaluate over $\bar{p}(x, y, z, u, \hat{x})$ in (3) or $p(x, y, z, u, \hat{x})$ in (4), and thus $(R_1, R_2, D, \triangle)$ is also in the achievable region. To do that, we show that the marginal distributions $\bar{p}(x, y, z, u)$ and $\bar{p}(x, u, \hat{x})$ induced by $\bar{p}(x, y, z, u, \hat{x})$ are equal to $p(x, y, z, u)$ and $p(x, u, \hat{x})$ induced by $p(x, y, z, u, \hat{x})$. By summing over $\hat{x}$ in (3) and (4), we have $\bar{p}(x, y, z, u) = p(x, y, z, u)$. To show that $\bar{p}(x, u, \hat{x}) = p(x, u, \hat{x})$, we consider $\bar{p}(x, u, \hat{x}) = \bar{p}(x, u)\bar{p}(\hat{x}|x, u)$. Note that, by summing over $(y, z, \hat{x})$ in (3) and (4), we get $\bar{p}(x, u) = p(x, u)$. Also, $p(\hat{x}|x, u) = \bar{p}(\hat{x}|x, u)$ since $p(\hat{x}|x, u)$ is the induced $\bar{p}(\hat{x}|x, u)$ by construction. Thus, we conclude that $\bar{p}(x, u, \hat{x}) = p(x, u, \hat{x})$. See also [3] for more details. For the bound on the cardinality of the set $\mathcal{U}$, it can be shown by using the support lemma [32] that it suffices that $\mathcal{U}$ should have $|\mathcal{Y}| - 1$ elements to preserve $P_Y$, plus four more for $H(Y|U), I(X; \hat{X}|U), H(X|U, Z)$, and the distortion constraint. ∎

*Remark 4:* For the logarithmic loss distortion case and the Gaussian source with quadratic distortion case specified before, it can be shown that the rate-distortion-leakage regions for the corresponding two-sided helper cases remain the same as those of the one-sided helper cases. This is a reminiscence of the well-known result in the Wyner-Ziv source coding problem with Gaussian source and quadratic distortion that the side information $Y^n$ at the encoder does not improve the rate-distortion function, i.e., $R_{X|Y}(D) = R^{WZ}(D) = \frac{1}{2}\log(\frac{\text{var}(X|Y)}{D})$ [34]. In our case, to prove the achievability, we simply neglect the coded side information at the encoder and achieve the same region as in the one-sided helper case. The converse proof also follows the one-sided helper case.

## IV. SECURE TRIANGULAR/CASCADE SOURCE CODING WITH A PUBLIC HELPER

In this section, we consider related problems where the data transmission involves an intermediate node, termed as *helper*. We assume that the communication through the helper is not secure, i.e., the helper itself is a public terminal to which we do not want to reveal too much information about the source sequence. We characterize the tradeoff between rate, distortion, and information leakage rate in the form of rate-distortion-leakage region for different settings of secure triangular/cascade source coding with a public helper (settings (A)-(D)) described earlier, see also Fig. 2). In our considered settings, the operation at the helper depends heavily on the side information available at the helper and the decoder. For example, if the side information at the helper is "degraded" with respect to that at the decoder, then the simple forwarding scheme is optimal. On the other hand, if the side information at the decoder is "degraded," it is optimal to perform decoding and re-encoding at the helper. Since the cascade settings are special cases of the triangular settings when removing the private link, i.e., setting $W_3$ to be constant, we only present the results and proofs for the triangular settings, and state the cascade results as corollaries.

## A. Triangular and Cascade Setting (A)

Setting (A) assumes that the side information $Y^n$ at a decoder is stronger than $Z^n$ at a helper in the sense that $X - Y - Z$ forms a Markov chain. We characterize the rate-distortion-leakage region of the triangular setting (A) (with the Markov chain assumption $X - Y - Z$) under logarithmic loss distortion measure, and for the Gaussian setting under quadratic distortion.

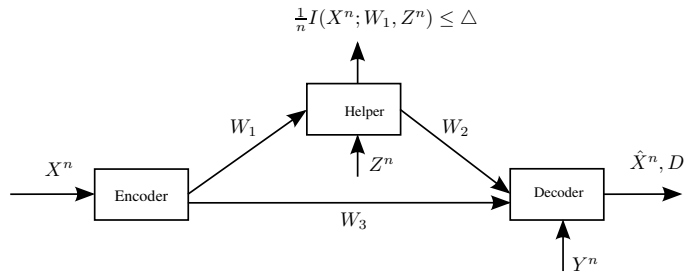

Fig. 5. Secure triangular source coding with a public helper, setting (A).

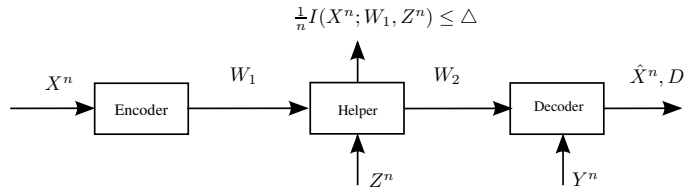

Fig. 6. Secure cascade source coding with a public helper, setting (A).

### 1) Logarithmic Loss Distortion:

*Theorem 5 (triangular (A), logarithmic loss):* The rate-distortion-leakage region $\mathcal{R}_{\text{tri(A), X-Y-Z, logloss}}$ under logarithmic loss distortion and $X - Y - Z$ assumption is the set of all tuples $(R_1, R_2, R_3, D, \triangle) \in \mathbb{R}_+^5$ that satisfy

$$R_1 \geq [H(X|Y) - D - R_3]^+,$$
$$R_2 \geq [H(X|Y) - D - R_3]^+,$$
$$\triangle \geq I(X;Z) + [H(X|Y) - D - R_3]^+. \tag{5}$$

*Remark 5:* Since we assume that $X - Y - Z$ forms a Markov chain, it is optimal to perform the Wyner-Ziv coding with $Y^n$ as side information at the receiver, and ignore the side information $Z^n$ by simply forwarding the index received at the helper. This results in the same rate constraints on $R_1$ and $R_2$. Moreover, with this forwarding scheme at hand, the rate-splitting of the index over the cascade and private links turns out to be optimal. Terms on the right hand side of the leakage rate constraint are simply due to the correlated side information $Z^n$, and the eavesdropped index at the helper.

*Proof of Theorem 5:*

*Sketch of Achievability*: The Wyner-Ziv coding at rate of $I(X;U|Y) + 2\epsilon = H(X|Y) - D + 2\epsilon$ is performed to satisfy the distortion constraint, where the equality is due to the choice of $U$ and the property of logarithmic loss distortion. If $H(X|Y) - D > R_3$, we perform rate-splitting on the Wyner-Ziv index. That is, we split the index into two parts, namely $w_1 \in [1:2^{n(H(X|Y)-D-R_3+\epsilon)}]$, and $w_3 \in [1:2^{n(R_3+\epsilon)}]$. The indices $w_1$ and $w_3$ are sent over the cascade link and the private (triangular) link, respectively. Then the helper forwards the index $w_1$ to the decoder. It can be seen that the rate and distortion constraints are satisfied. As for the analysis of leakage rate, we consider the normalized mutual information averaged over all codebooks $\mathcal{C}_n$,

$$I(X^n; W_1, Z^n|\mathcal{C}_n)$$
$$= I(X^n; Z^n|\mathcal{C}_n) + I(X^n; W_1|Z^n, \mathcal{C}_n)$$
$$\leq I(X^n; Z^n|\mathcal{C}_n) + H(W_1|Z^n, \mathcal{C}_n)$$
$$\overset{(a)}{\leq} n[I(X;Z) + H(X|Y) - D - R_3 + \epsilon]$$

where $(a)$ follows from the facts that $(X^n, Z^n)$ are i.i.d. and independent of the codebook, and from the codebook generation that we have $W_1 \in [1:2^{n(H(X|Y)-D-R_3+\epsilon)}]$.

On the other hand, if $H(X|Y) - D < R_3$, we send the Wyner-Ziv index over the private link, and send nothing over the cascade links, i.e., $R_1 \geq 0, R_2 \geq 0$ are achievable. The corresponding leakage rate is $\frac{1}{n}I(X^n; W_1, Z^n|\mathcal{C}_n) = \frac{1}{n}I(X^n; Z^n) = I(X; Z)$. The converse proof is given in Appendix F. ∎

*Corollary 2 (cascade (A), logarithmic loss):* The rate-distortion-leakage region $\mathcal{R}_{\text{cas(A), X-Y-Z, logloss}}$ under logarithmic loss distortion and $X - Y - Z$ assumption is the set of all tuples $(R_1, R_2, D, \triangle) \in \mathbb{R}^4_+$ that satisfy

$$R_1 \geq [H(X|Y) - D]^+,$$
$$R_2 \geq [H(X|Y) - D]^+,$$
$$\triangle \geq I(X;Z) + [H(X|Y) - D]^+.$$

*2) Gaussian Source with Quadratic Distortion and Markov Chain relation $X - Y - Z$:* Let the sequences $(X^n, Y^n, Z^n)$ be i.i.d. according to $P_{X,Y,Z}$. We assume that $X$ has a Gaussian distribution with zero mean and variance $\sigma_X^2$, i.e., $X \sim \mathcal{N}(0, \sigma_X^2)$. Let $Y = X + N_1, N_1 \sim \mathcal{N}(0, \sigma_{N_1}^2)$ independent of $X$, and $Z = Y + N_2, N_2 \sim \mathcal{N}(0, \sigma_{N_2}^2)$ independent of $(X, Y, N_1)$, where $\sigma_X^2, \sigma_{N_1}^2, \sigma_{N_2}^2 > 0$. This satisfies the Markov assumption $X - Y - Z$.

*Theorem 6 (triangular (A), Gaussian):* The rate-distortion-leakage region for a Gaussian source with quadratic distortion under the Markov assumption $X - Y - Z$, $\mathcal{R}_{\text{tri(A), X-Y-Z, Gaussian}}$, is the set of all tuples $(R_1, R_2, R_3, D, \triangle) \in$

$\mathbb{R}^5_+$ that satisfy

$$R_1 \geq [\frac{1}{2} \log (\sigma^2/D) - R_3]^+,$$
$$R_2 \geq [\frac{1}{2} \log (\sigma^2/D) - R_3]^+,$$
$$\triangle \geq \frac{1}{2} \log \left(1 + \frac{\sigma_X^2}{\sigma_{N_1}^2 + \sigma_{N_2}^2}\right) + [\frac{1}{2} \log (\sigma^2/D) - R_3]^+,$$

where $\sigma^2 = \frac{\sigma_X^2 \sigma_{N_1}^2}{\sigma_X^2 + \sigma_{N_1}^2}$.

*Proof:* The proof is given in Appendix G. ∎

*Corollary 3 (cascade (A), Gaussian):* The rate-distortion-leakage region for a Gaussian source with quadratic distortion under the Markov assumption $X - Y - Z$, $\mathcal{R}_{\text{cas(A), X-Y-Z, Gaussian}}$, is the set of all tuples $(R_1, R_2, D, \triangle) \in \mathbb{R}^4_+$ that satisfy

$$R_1 \geq \frac{1}{2} \log (\sigma^2/D),$$
$$R_2 \geq \frac{1}{2} \log (\sigma^2/D),$$
$$\triangle \geq \frac{1}{2} \log \left(1 + \frac{\sigma_X^2}{\sigma_{N_1}^2 + \sigma_{N_2}^2}\right) + \frac{1}{2} \log (\sigma^2/D),$$

where $\sigma^2 = \frac{\sigma_X^2 \sigma_{N_1}^2}{\sigma_X^2 + \sigma_{N_1}^2}$.

Based on the triangular setting in Fig. 5, one might consider a related scenario where the encoder can only "broadcast" (BC) the same source description to the helper and the decoder over the rate-limited digital links, in the sense of Fig. 3, i.e., $W_3 = W_1$. Based on the source description and some side information, the helper generates a new description and sends it to the decoder. The rest of the problem formulation of this triangular setting is similar to that of Fig. 5. We characterize the rate-distortion-leakage region under the logarithmic loss distortion and the Markov assumption $X - Y - Z$.

*Theorem 7 (triangular (A), logarithmic loss, BC):* The rate-distortion-leakage region $\mathcal{R}_{\text{tri(A), X-Y-Z, logloss, BC}}$ under logarithmic loss distortion is the set of all tuples $(R_1, R_2, D, \triangle) \in \mathbb{R}^4_+$ that satisfy

$$R_1 \geq [H(X|Y) - D]^+,$$
$$R_2 \geq 0,$$
$$\triangle \geq I(X;Z) + [H(X|Y) - D]^+.$$

*Proof:* If $H(X|Y) - D > 0$, the achievability proof follows the Wyner-Ziv coding for the encoder/decoder pair at rate above $I(X;U|Y) = H(X|Y) - D$. Since the index $W_1$ is also available at the decoder, the helper does not need to send anything to the decoder (due to data processing theorem). The leakage proof follows similarly as the proof of Theorem 5. On the other hand, if $H(X|Y) - D < 0$, the encoder does not need to send anything, i.e., $R_1 \geq 0, R_2 \geq 0$ are achievable. The corresponding leakage rate is $\frac{1}{n} I(X^n; W_1, Z^n | \mathcal{C}_n) = \frac{1}{n} I(X^n; Z^n) = I(X;Z)$. Converse proofs for $R_1$ and $\triangle$ constraints follow similarly as the proof of Theorem 5, while $R_2 \geq 0$ is trivial. ∎

*Remark 6:* In this case, the helper does not help to provide any additional information to the decoder due to the Markov relation $X^n - Y^n - Z^n$. That is, given $(W_1, Y^n)$, the decoder has already all information about the source available. Note also that this setting is similar to the source coding setting considered in [5] with the cooperation link from the helper to the decoder. However, the cooperation link does not provide any extra information to the decoder.

## B. Triangular and Cascade setting (B)

Setting (B) assumes that the common side information $Y^n$ is available at both encoder and decoder. This allows the encoder and decoder a possibility to generate a secret key for protecting the source description sent through the public helper. We characterize the rate-distortion-leakage region of the triangular setting (B) (with the Markov chain assumption $X - Y - Z$) under logarithmic loss distortion measure, and for the Gaussian setting under quadratic distortion.

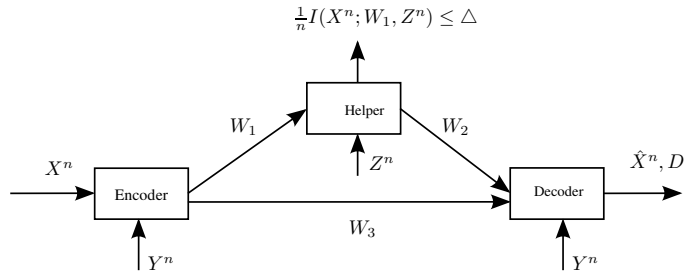

Fig. 7. Secure triangular source coding with a public helper, setting (B).

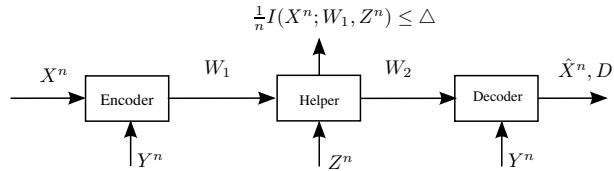

Fig. 8. Secure cascade source coding with a public helper, setting (B).

*1) Logarithmic Loss Distortion:*

*Theorem 8 (triangular (B), logloss):* The rate-distortion-leakage region $\mathcal{R}_{\text{tri(B)}, \text{X-Y-Z}, \text{logloss}}$ under logarithmic loss distortion is the set of all tuples $(R_1, R_2, R_3, D, \triangle) \in \mathbb{R}_+^5$ that satisfy

$$R_1 \geq [H(X|Y) - D - R_3]^+, \tag{6a}$$

$$R_2 \geq [H(X|Y) - D - R_3]^+, \tag{6b}$$

$$\triangle \geq I(X;Z) + [H(X|Y) - D - R_3 - H(Y|X,Z)]^+. \tag{6c}$$

*Remark 7:* We first note that the availability of side information $Y^n$ at the encoder does not improve the rate-distortion tradeoff under a logarithmic loss distortion, with respect to the Wyner-Ziv setting [8] (like in the Gaussian case [34]). Interestingly though, the common side information at the encoder helps to reduce the leakage rate at the helper by allowing the encoder and the decoder to generate a secret key. We can see this from the leakage constraint (6c) above where the leakage rate consists of contributions from the eavesdropper's side information $I(X; Z)$ and from the eavesdropped source description which is partially protected by the secret key of rate $\min\{H(Y|X,Z), H(X|Y) - D - R_3\}$ (cf. (5) in Theorem 5 where there is no leakage reduction from the secret key). This role of side information at the encoder and the decoder in another secure source coding setting is also studied in [33].

*Proof of Theorem 8:*

*Sketch of Achievability*: The proof follows similarly as in previous triangular case with the additional steps of secret key generation using $y^n$. The achievable scheme below is also similar to that found in [33]. That is, the Wyner-Ziv coding at rate $I(X; U|Y) + 2\epsilon = H(X|Y) - D + 2\epsilon$ is performed to satisfy the distortion constraint. Then we perform rate-splitting on the Wyner-Ziv index by splitting it into two parts, namely $w_1 \in [1 : 2^{n(H(X|Y)-D-R_3+\epsilon)}]$, and $w_3 \in [1 : 2^{n(R_3+\epsilon)}]$. Next we distinguish between two cases where we further split the index $w_1$ and where the key rate is sufficient for scrambling the whole index $w_1$.

If $H(X|Y) - D - R_3 > H(Y|X, Z)$, we further split $w_1$ into $w_{11} \in [1 : 2^{n(H(X|Y)-D-R_3-H(Y|X,Z)+\epsilon)}]$ and $w_{12} \in [1 : 2^{nH(Y|X,Z)}]$. Then the secret key $k$ is generated by randomly and independently partitioning sequences in $\mathcal{Y}^n$ into $2^{nH(Y|X,Z)}$ bins and choosing $k$ as the corresponding bin index of the given $y^n$. The encoder sends $w_{11}$ and $w_{12} \oplus k$ over the cascade link, and $w_3$ over the private link, where $w_{12} \oplus k$ denotes the modulo operation, $(w_{12} + k) \mod 2^{nH(Y|X,Z)}$[1]. The helper forwards the index $w_{11}$ and $w_{12} \oplus k$ to the decoder. The decoder can recover $w_{12}$ from its key generated by $y^n$. We can show that the tuples satisfying (6) where $[a]^+ = a$ in (6c) are achievable.

If $H(X|Y) - D - R_3 < H(Y|X, Z)$, the secret key is generated by randomly and independently partitioning sequences in $\mathcal{Y}^n$ into $2^{n(H(X|Y)-D-R_3+\epsilon)}$ bins and choosing the corresponding bin index of given $y^n$ as a key. The encoder sends $w_1 \oplus k$ over the cascade link, and $w_3$ over the private link. The helper forwards $w_1 \oplus k$ to the decoder. We can show that the tuples satisfying (6) where $[a]^+ = 0$ in (6c) are achievable. For the detailed achievability proof and converse proof, please see Appendix H and I. ∎

*Corollary 4 (cascade (B), logloss):* The rate-distortion-leakage region $\mathcal{R}_{\text{cas(B), X-Y-Z, logloss}}$ under logarithmic loss distortion is the set of all tuples $(R_1, R_2, D, \triangle) \in \mathbb{R}_+^4$ that satisfy

$$R_1 \geq [H(X|Y) - D]^+,$$

$$R_2 \geq [H(X|Y) - D]^+,$$

$$\triangle \geq I(X; Z) + [H(X|Y) - D - H(Y|X, Z)]^+.$$

---

[1] Here, we have $w_{12}, k \in [1 : 2^{nH(Y|X,Z)}]$. Thus, in the modulo operation, 0 is mapped to $2^{nH(Y|X,Z)}$.

*2) Gaussian Source with Quadratic Distortion and Markov Chain relation $X - Y - Z$:* Let the sequences $(X^n, Y^n, Z^n)$ be i.i.d. according to $P_{X,Y,Z}$. We assume that $X$ has a Gaussian distribution with zero mean and variance $\sigma_X^2$, i.e., $X \sim \mathcal{N}(0, \sigma_X^2)$. Let $Y = X + N_1, N_1 \sim \mathcal{N}(0, \sigma_{N_1}^2)$ independent of $X$, and $Z = Y + N_2, N_2 \sim \mathcal{N}(0, \sigma_{N_2}^2)$ independent of $(X, Y, N_1)$, where $\sigma_X^2, \sigma_{N_1}^2, \sigma_{N_2}^2 > 0$. This satisfies the Markov assumption $X - Y - Z$.

*Theorem 9 (triangular (B), Gaussian):* The rate-distortion-leakage region for a Gaussian source with quadratic distortion under the Markov assumption $X - Y - Z$, $\mathcal{R}_{\text{tri(B), X-Y-Z, Gaussian}}$, is the set of all tuples $(R_1, R_2, R_3, D, \triangle) \in \mathbb{R}_+^5$ that satisfy

$$R_1 \geq [\frac{1}{2} \log(\sigma^2/D) - R_3]^+,$$

$$R_2 \geq [\frac{1}{2} \log(\sigma^2/D) - R_3]^+,$$

$$\triangle \geq \frac{1}{2} \log\left(1 + \frac{\sigma_X^2}{\sigma_{N_1}^2 + \sigma_{N_2}^2}\right).$$

*Proof:* The availability of Gaussian side information $Y^n$ at the encoder and decoder allows us to generate a *discrete* secret key at arbitrarily high rate. This implies that we can essentially protect the whole source description sent over the rate limited link to the helper, and the only leakage to the eavesdropper is due to the eavesdropper's correlated side information $Z^n$. The detailed proof is given in Appendix J. ∎

*Corollary 5 (cascade (B), Gaussian):* The rate-distortion-leakage region for a Gaussian source with quadratic distortion under the Markov assumption $X - Y - Z$, $\mathcal{R}_{\text{cas(B), X-Y-Z, Gaussian}}$, is the set of all tuples $(R_1, R_2, D, \triangle) \in \mathbb{R}_+^4$ that satisfy

$$R_1 \geq \frac{1}{2} \log(\sigma^2/D),$$

$$R_2 \geq \frac{1}{2} \log(\sigma^2/D),$$

$$\triangle \geq \frac{1}{2} \log\left(1 + \frac{\sigma_X^2}{\sigma_{N_1}^2 + \sigma_{N_2}^2}\right).$$

Again, based on the triangular setting in Fig. 7, one can consider a related scenario where the encoder can only "broadcast" the same source description to the helper and the decoder over the rate-limited digital link, in the sense of Fig. 3. We characterize the rate-distortion-leakage region under the logarithmic loss distortion. Similarly to the result in Theorem 7, the helper is not helpful in terms of helping the transmission due to the Markov assumption $X - Y - Z$. In other words, $W_2$ does not provide any extra information to the decoder.

*Theorem 10 (triangular (B), logarithmic loss, BC):* The rate-distortion-leakage region $\mathcal{R}_{\text{tri(B), X-Y-Z, logloss, BC}}$ under logarithmic loss distortion is the set of all tuples $(R_1, R_2, D, \triangle) \in \mathbb{R}_+^4$ that satisfy

$$R_1 \geq [H(X|Y) - D]^+,$$

$$R_2 \geq 0,$$

$$\triangle \geq I(X; Z) + [H(X|Y) - D - H(Y|X, Z)]^+.$$

*Proof:* Since we have the Markov assumption $X - Y - Z$ and the index $W_1$ is also available at the decoder, the helper does not need to send anything to the decoder. Hence, the problem turns into a standard secure source

coding with side information at both encoder and decoder. The proof follows similarly as that of Theorem 8. ∎

## C. Triangular and Cascade Setting (C)

Setting (C) assumes that the helper has no side information. We characterize the rate-distortion-leakage region for the triangular/cascade setting (C) under general distortion.

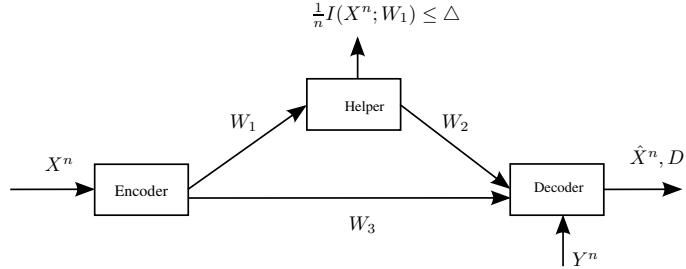

Fig. 9. Secure triangular source coding with a public helper, setting (C).

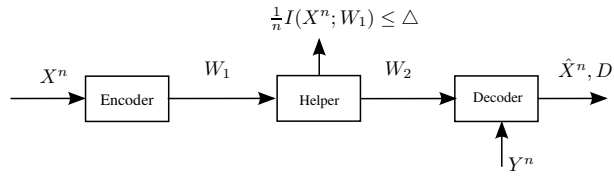

Fig. 10. Secure cascade source coding with a public helper, setting (C).

*Theorem 11 (triangular (C)):* The rate-distortion-leakage region $\mathcal{R}_{\text{tri(C)}}$ is the set of all tuples $(R_1, R_2, R_3, D, \triangle) \in \mathbb{R}_+^5$ for which there exist a random variable $U \in \mathcal{U}$ such that $U - X - Y$ forms a Markov chain, and a function $g : \mathcal{U} \times \mathcal{Y} \to \hat{\mathcal{X}}$ that satisfy

$$R_1 \geq [I(X;U|Y) - R_3]^+,$$

$$R_2 \geq [I(X;U|Y) - R_3]^+,$$

$$D \geq E[d(X, g(U,Y))],$$

$$\triangle \geq [I(X;U|Y) - R_3]^+.$$

The cardinality of the alphabet of the auxiliary random variable can be upperbounded as $|\mathcal{U}| \leq |\mathcal{X}| + 1$.

*Remark 8:* Since there is no side information at the helper, it is obvious that the optimal scheme at the helper is to simply forward the source description, i.e., setting $W_2 = W_1$. In this case, unlike setting (A) in Fig. 5 and 6, we are able to solve the problem under a general distortion measure since the problem essentially reduces to the Wyner-Ziv problem with an additional leakage rate constraint.

*Proof of Theorem 11:*

*Sketch of Achievability*: The proof is similar to that of triangular setting (A) where we use rate splitting. The Wyner-Ziv coding at rate of $I(X;U|Y)+2\epsilon$ is performed to satisfy the distortion constraint. Then we perform rate-splitting on the Wyner-Ziv index. That is, we split the index into two parts, namely $w_1 \in [1, 2^{n(I(X;U|Y)-R_3+\epsilon)}]$, and $w_3 \in [1, 2^{n(R_3+\epsilon)}]$. The indices $w_1$ and $w_3$ are sent over the cascade link and the private (triangular) link, respectively. The helper forwards the index $w_1$ to the decoder. The analysis of distortion follows from the analysis for the Wyner-Ziv setting in [35, Ch. 11]. As for the analysis of leakage rate, we consider the normalized mutual information averaged over all codebooks,

$$I(X^n; W_1|\mathcal{C}_n) \leq H(W_1|\mathcal{C}_n) \stackrel{(a)}{\leq} n[I(X;U|Y) - R_3 + \epsilon]$$

where $(a)$ follows from the codebook generation that we have $W_1 \in [1 : 2^{n(I(X;U|Y)-R_3+\epsilon)}]$.

The converse proof follows similarly as in the triangular setting (A) and is given in Appendix K. ∎

*Corollary 6 (cascade (C)):* The rate-distortion-leakage region $\mathcal{R}_{\text{cas(C)}}$ is the set of all tuples $(R_1, R_2, D, \triangle) \in \mathbb{R}_+^4$ for which there exist a random variable $U \in \mathcal{U}$ such that $U - X - Y$ forms a Markov chain, and a function $g : \mathcal{U} \times \mathcal{Y} \to \hat{\mathcal{X}}$ that satisfy

$$R_1 \geq I(X;U|Y),$$
$$R_2 \geq I(X;U|Y),$$
$$D \geq E[d(X, g(U,Y))],$$
$$\triangle \geq I(X;U|Y).$$

The cardinality of the auxiliary random variable can be upperbounded as $|\mathcal{U}| \leq |\mathcal{X}| + 1$.

As before, based on the triangular setting in Fig. 9, one can consider a related scenario where the encoder broadcasts the same source description to the helper and the decoder, in the sense of Fig. 3. We characterize the rate-distortion-leakage region for a general distortion. Similarly to the result in Theorem 7 and 10, the helper is not helpful in terms of providing additional information to the decoder.

*Theorem 12 (triangular (C), BC):* The rate-distortion-leakage region $\mathcal{R}_{\text{tri(C), BC}}$ is the set of all tuples $(R_1, R_2, D, \triangle) \in \mathbb{R}_+^4$ for which there exist a random variable $U \in \mathcal{U}$ such that $U - X - Y$ forms a Markov chain, and a function $g : \mathcal{U} \times \mathcal{Y} \to \hat{\mathcal{X}}$ that satisfy

$$R_1 \geq I(X;U|Y),$$
$$R_2 \geq 0,$$
$$D \geq E[d(X, g(U,Y))],$$
$$\triangle \geq I(X;U|Y).$$

The cardinality of the auxiliary random variable can be upperbounded as $|\mathcal{U}| \leq |\mathcal{X}| + 1$.

*Proof:* Since the index $W_1$ is also available at the decoder, the helper does not need to send anything to the decoder. The proof essentially follows the Wyner-Ziv coding proof [35, Ch. 11] and the proof of information leakage constraint follows similarly as in Theorem 11. ∎

## D. Triangular and Cascade Setting (D)

In setting (D), we consider the case where side information $Z^n$ at the helper is assumed to be available to the encoder, as depicted in Fig. 11 and 12, under the Markov assumption $X - Z - Y$. This setting is "dual" to the setting (B) in the sense that we switch the order of side information degradedness and the availability of helper's side information or decoder's side information at the encoder. We characterize the rate-distortion-leakage region for triangular/cascade setting (D) under general distortion.

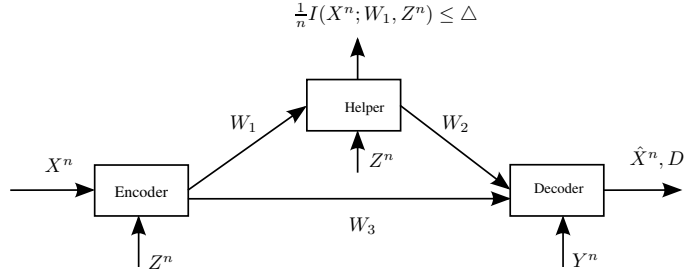

Fig. 11. Secure triangular source coding with a public helper with $X - Z - Y$, setting (D).

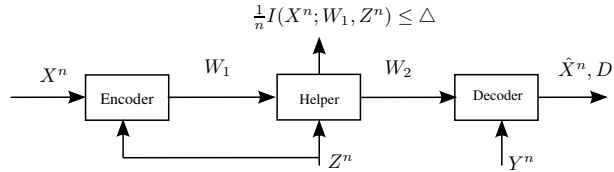

Fig. 12. Secure cascade source coding with a public helper, $X - Z - Y$, setting (D).

*Theorem 13 (triangular (D)):* The rate-distortion-leakage region $\mathcal{R}_{\text{tri(D)}, \text{X-Z-Y}}$ is the set of all tuples $(R_1, R_2, R_3, D, \triangle) \in \mathbb{R}_+^5$ for which there exist random variables $U \in \mathcal{U}$ and $V \in \mathcal{V}$ such that $(U, V, X) - Z - Y$ forms a Markov chain, and a function $g : \mathcal{U} \times \mathcal{V} \times \mathcal{Y} \to \hat{\mathcal{X}}$ that satisfy

$$R_1 \geq I(X; U|Z),$$
$$R_2 \geq I(X, Z; U|Y),$$
$$R_3 \geq I(X, Z; V|U, Y),$$
$$D \geq E[d(X, g(U, V, Y))],$$
$$\triangle \geq I(X; U, Z).$$

The cardinalities of the alphabets of the auxiliary random variables can be upperbounded as $|\mathcal{U}| \leq |\mathcal{X}||\mathcal{Z}| + 3$ and $|\mathcal{V}| \leq (|\mathcal{X}||\mathcal{Z}| + 3)(|\mathcal{X}||\mathcal{Z}| + 1)$.

*Remark 9:* Since we assume a new order of side information degradedness $X - Z - Y$, the optimal scheme at the helper becomes *decode and re-bin*. In other words, the side information $Z^n$ at the helper is useful in providing extra information to the decoder. We note that if the leakage constraint at the helper is replaced by the decoding constraint under some distortion, the problem turns into the original triangular/cascade source coding problem by Chia et al. [15].

*Proof of Theorem 13:*

*Sketch of Achievability*: The achievable scheme follows the *decode and re-bin* scheme of [15]. That is, randomly generate $2^{n(I(X,Z;U)+\epsilon)}$ sequences $u^n(\tilde{w}_1) \sim \prod_{i=1}^n P_U(u_i(\tilde{w}_1))$, $\tilde{w}_1 \in [1 : 2^{n(I(X,Z;U)+\epsilon)}]$. Then distribute these sequences uniformly into $2^{n(I(X;U|Z)+2\epsilon)}$ bins $b_{u_1}(w_1)$, $w_1 \in [1 : 2^{n(I(X;U|Z)+2\epsilon)}]$. In addition, we distribute them also uniformly into another $2^{n(I(X,Z;U|Y)+2\epsilon)}$ bins $b_{u_2}(w_2)$, $w_2 \in [1 : 2^{n(I(X,Z;U|Y)+2\epsilon)}]$. Furthermore, for each $\tilde{w}_1$, randomly generate $2^{n(I(X,Z;V|U)+\epsilon)}$ sequences $v^n(\tilde{w}_1, \tilde{w}_3) \sim \prod_{i=1}^n P_{V|U}(\cdot|u_i(\tilde{w}_1))$, and distribute them uniformly into $2^{n(I(X,Z;V|U,Y)+2\epsilon)}$ bins $b_v(w_3)$, $w_3 \in [1 : 2^{n(I(X,Z;V|U,Y)+2\epsilon)}]$. For encoding, the encoder looks for a sequence $u^n$ that is jointly typical with $(x^n, z^n)$. If there is more than one such sequence, it selects one of them uniformly at random. If there is no such $u^n$, it selects one out of $2^{n(I(X,Z;U)+\epsilon)}$ uniformly at random. With high probability, there exists such $u^n$ since there are $2^{n(I(X,Z;U)+\epsilon)}$ codewords generated. Then it transmits the corresponding bin index $w_1$ to the helper. Also, the encoder looks for $v^n$ that is jointly typical with $(x^n, z^n, u^n)$. If there is more than one, it selects one of them uniformly at random. If there is no such $v^n$, it selects one out of $2^{n(I(X,Z;V|U)+\epsilon)}$ uniformly at random. With high probability, there exists such $v^n$ since there are $2^{n(I(X,Z;V|U)+\epsilon)}$ codewords generated. Then it transmits the corresponding bin index $w_3$ to the decoder over the private link. Upon receiving the bin index $w_1$, the helper node looks for the unique $u^n$ such that it is jointly typical with the side information $z^n$. With high probability, it will find the unique and correct one since there are $2^{n(I(U;Z)-\epsilon)}$ codewords in each bin $b_{u_1}(w_1)$. After that the helper looks for the corresponding bin $b_{u_2}(w_2)$ such that the decoded $u^n \in b_{u_2}(w_2)$, and transmit the bin index $w_2$ to the decoder. The decoder, with high probability, will successively find the unique and correct $u^n$ and $v^n$ that are jointly typical with $y^n$ since there are $2^{n(I(U;Y)-\epsilon)}$ codewords in each bin $b_{u_2}(w_2)$, and there are $2^{n(I(V;Y|U)-\epsilon)}$ codewords in each bin $b_v(w_3)$. Then $\hat{x}^n$ is put out as a source reconstruction, where $\hat{x}_i = g(u_i, v_i, y_i)$. Since $(x^n, u^n, v^n, y^n)$ are jointly typical, we can show that $D \geq E[d(X, g(U, V, Y))]$ is achievable.

As for the analysis of leakage rate, we consider the normalized mutual information averaged over all codebooks,

$$I(X^n; W_1, Z^n | \mathcal{C}_n)$$
$$= I(X^n; Z^n | \mathcal{C}_n) + I(X^n; W_1 | Z^n, \mathcal{C}_n)$$
$$\leq I(X^n; Z^n | \mathcal{C}_n) + H(W_1 | Z^n, \mathcal{C}_n)$$
$$\overset{(a)}{\leq} n[I(X; Z) + I(X; U|Z) + \delta_\epsilon]$$
$$= n[I(X; U, Z) + \delta_\epsilon]$$

where $(a)$ follows from the facts that $(X^n, Z^n)$ are i.i.d. and independent of the codebook, and from the codebook generation that we have $W_1 \in [1 : 2^{n(I(X;U|Z)+\epsilon)}]$. The converse proof is given in Appendix L. ■

*Corollary 7 (cascade (D)):* The rate-distortion-leakage region $\mathcal{R}_{\text{cas(D), X-Z-Y}}$ is the set of all tuples $(R_1, R_2, D, \triangle) \in \mathbb{R}_+^4$ for which there exist a random variable $U \in \mathcal{U}$ such that $(U, X) - Z - Y$ forms a Markov chain, and a function

$g : \mathcal{U} \times \mathcal{Y} \to \hat{\mathcal{X}}$ that satisfy

$$R_1 \geq I(X; U|Z),$$

$$R_2 \geq I(X, Z; U|Y),$$

$$D \geq E[d(X, g(U, Y))],$$

$$\triangle \geq I(X; U, Z).$$

The cardinality of the alphabet of the auxiliary random variable can be upperbounded as $|\mathcal{U}| \leq |\mathcal{X}||\mathcal{Z}| + 2$.

We also consider the scenario where the encoder broadcasts the source description to the helper and the decoder, in the sense of Fig. 3. We characterize the rate-distortion-leakage region for a general distortion. In this case, unlike the previous three cases under settings (A)-(C), the helper is useful in terms of supporting the transmission since its side information $Z^n$ is "stronger" than $Y^n$ at the decoder due to the assumption that $X - Z - Y$ forms a Markov chain. To satisfy the distortion constraint, we are required to satisfy both constraints on the individual rate $R_1$ and sum-rate $R_1 + R_2$ (cf. successive refinement problem).

*Theorem 14 (triangular (D), BC):* The rate-distortion-leakage region $\mathcal{R}_{\text{tri(D), X-Z-Y, BC}}$ is the set of all tuples $(R_1, R_2, D, \triangle)$ for which there exist a random variable $U \in \mathcal{U}$ such that $(U, X) - Z - Y$ forms a Markov chain, and a function $g : \mathcal{U} \times \mathcal{Y} \to \hat{\mathcal{X}}$ that satisfy

$$R_1 \geq I(X; U|Z),$$

$$R_1 + R_2 \geq I(X, Z; U|Y),$$

$$D \geq E[d(X, g(U, Y))],$$

$$\triangle \geq I(X; U, Z).$$

The cardinality of the auxiliary random variable can be upperbounded as $|\mathcal{U}| \leq |\mathcal{X}||\mathcal{Z}| + 2$.

*Proof:* Since both indices $W_1$ and $W_2$ are available at the decoder, we need to satisfy the sum-rate constraint $R_1 + R_2$ instead of the individual rate $R_2$. For achievability, the encoder sends the partial bin index of the selected codeword at rate $R_1$ and then the helper performs *decode and re-bin* and sends the rest of the bin index at rate $R_2$ to the decoder. The detailed proof is given in Appendix M. ∎

*Remark 10:* Here we discuss the optimal operation at the helper in all considered settings. Since the private link in the triangular setting can only provide additional information subject to its rate constraint, the processing ambiguity lies only in the cascade transmission, i.e., what is the best relaying strategy at the helper? For ease of discussion, we will for now neglect the private link, and argue that when the side information at the helper is degraded with respect to that at the decoder, forwarding scheme at the helper is optimal; otherwise, it is optimal to employ a decode-and-re encode type scheme.

- Let us consider the setting (A) in which we assume that $X - Y - Z$ forms a Markov chain (the discussion for setting (B) and (C) follows similarly). On the cascade link, in order to attain low distortion at the decoder, we wish to compress the source so that the decoder, upon receiving $(W_2, Y^n)$, can extract as much information about the source as possible, i.e., maximizing $I(X^n; W_2, Y^n)$. The joint pmf of this setting after summing out the reconstruction sequences is given by $P_{X^n,Y^n} P_{Z^n|Y^n} P_{W_1|X^n} P_{W_2|W_1,Z^n}$. Data processing inequality implies that $I(X^n; W_2, Y^n) \leq I(X^n; W_1, Y^n)$. This suggests that the forwarding scheme at the helper (setting $W_2 = W_1$) is a good strategy for this setting, and it is in fact optimal in this case.

- On the other case (setting (D)) where we assume that $X - Z - Y$ forms a Markov chain, the joint pmf after summing out the reconstruction sequences is given by $P_{X^n,Z^n} P_{Y^n|Z^n} P_{W_1|X^n,Z^n} P_{W_2|W_1,Z^n}$. To see if the forwarding scheme is still optimal, we consider the following inequality (derived from the joint pmf using the data processing inequality), $I(X^n; W_2, Y^n) \leq I(X^n; W_1, Z^n)$. The inequality suggests that, based on information available, the helper can extract more information about $X^n$ than the decoder does, regardless of what the helper scheme is. Since $W_2$ is generated based on $(W_1, Z^n)$, it is reasonable that the helper takes into account the knowledge about $Z^n$ in relaying the information, rather than just forwarding $W_1$. It turns out that the decode-and-re encode type scheme is optimal in this case.

## V. CONCLUSION

We study secure source coding problems with a public helper that supports the transmission while there is a risk for information leakage. Two classes of problems are considered, namely secure source coding with a helper where the helper link is eavesdropped, and secure triangular/cascade source coding with a public helper who is friendly but curious. We are interested in how the helper can facilitate transmission in these unsecured scenarios. We characterize the rate-distortion-leakage regions for different settings. In the first class of the problems, we present the rate-distortion-leakage regions for one-sided and two-sided helper cases under some specific distortion measure, and show that a standard coding scheme is optimal. We found that, for the logarithmic loss distortion case and the case of a Gaussian source with quadratic distortion under a Markov relation, the region is the same for both the one-sided and two-sided settings. This observation provides evidence that the availability of (coded) side information at the encoder does not improve the rate-distortion-leakage tradeoff. Furthermore, in triangular/cascade settings, we solve several special cases and observe that the optimal operation at the helper in our coding scheme depends heavily on the order of side information degradedness, i.e., when $X - Y - Z$ forms a Markov chain, the forwarding scheme is optimal, and when $X - Z - Y$ forms a Markov chain, the *decode and re-bin* scheme is optimal. The scenario where the common side information is available at both encoder and decoder, but not at the helper, gives rise to a new, interesting scheme involving secret key generation, and also reveals some connection to secure network coding. It is interesting to note that, in some cases, the availability of side information at the encoder is not useful in terms of rate-distortion tradeoff, but it can be used for secret key generation and thus helps to improve the leakage rate.

## ACKNOWLEDGEMENT

The authors wish to thank Rajiv Soundararajan for helpful discussions on the cascade and triangular source coding setup with equivocation constraints. This work was supported in part by the Swedish Research Council and the NSF Center for Science of Information (CSoI).

# APPENDIX A
## PROOF OF LEMMA 1

By definition of the reconstruction function, we get $g(u) \triangleq \hat{x}(x|u) \triangleq \Pr(X = x|U = u)$. Then we obtain
$E[d(X, g(U))] = \sum_{x \in \mathcal{X}, u \in \mathcal{U}} p(x, u) d(x, \hat{x}(\cdot|u)) = \sum_{x \in \mathcal{X}, u \in \mathcal{U}} p(x, u) \log(\frac{1}{p(x|u)}) = H(X|U)$.

# APPENDIX B
## PROOF OF LEMMA 2

By definition of the reconstruction alphabet, we consider the reconstruction $\hat{X}^n$ to be a probability distribution on $\mathcal{X}^n$ conditioned on $Z$. In particular, if $\hat{x}^n = g^{(n)}(z)$, we define $s(x^n|z) = \prod_{i=1}^{n} \hat{x}_i(x_i|z)$. Note that $s$ is a probability measure on $\mathcal{X}^n$. We obtain the following bound on the expected distortion conditioned on $Z = z$,

$$
\begin{aligned}
E[d^{(n)}(X^n, g^{(n)}(z))|Z = z] &= E[\frac{1}{n}\sum_{i=1}^{n} d(X_i, g_i^{(n)}(z))|Z = z] \\
&= \sum_{x^n \in \mathcal{X}^n} p(x^n|z) \frac{1}{n} \sum_{i=1}^{n} d(x_i, g_i^{(n)}(z)) \\
&= \sum_{x^n \in \mathcal{X}^n} p(x^n|z) \frac{1}{n} \sum_{i=1}^{n} \log(\frac{1}{\hat{x}_i(x_i|z)}) \\
&= \frac{1}{n} \sum_{x^n \in \mathcal{X}^n} p(x^n|z) \log(\frac{1}{s(x^n|z)}) \\
&= \frac{1}{n} \sum_{x^n \in \mathcal{X}^n} p(x^n|z) \log(\frac{p(x^n|z)}{s(x^n|z)} \cdot \frac{1}{p(x^n|z)}) \\
&= \frac{1}{n} \sum_{x^n \in \mathcal{X}^n} p(x^n|z) \log(\frac{p(x^n|z)}{s(x^n|z)}) + \frac{1}{n} \sum_{x^n \in \mathcal{X}^n} p(x^n|z) \log(\frac{1}{p(x^n|z)}) \\
&= \frac{1}{n} D(p(x^n|z) \| s(x^n|z)) + \frac{1}{n} H(X^n|Z = z) \\
&\geq \frac{1}{n} H(X^n|Z = z).
\end{aligned}
$$

By averaging both sides over all $z \in \mathcal{Z}$, from the law of total expectation, we obtain the desired result.

## APPENDIX C

### PROOF OF LEMMA 3

Consider the term $H(Y^n|U^n, X^n, Z^n, \mathcal{C}_n)$. Let us define $T$ to be a binary random variable taking value 1 if $(Y^n, U^n, X^n, Z^n)$ are jointly typical, and 0 otherwise. If $\Pr(T=0) \leq \delta_\epsilon$ for $n$ sufficiently large, we get

$$H(Y^n|W_2, X^n, Z^n, \mathcal{C}_n)$$
$$\leq H(Y^n|U^n, X^n, Z^n, \mathcal{C}_n)$$
$$\leq H(Y^n|U^n, X^n, Z^n, T) + H(T)$$
$$\leq \Pr(T=0) \cdot H(Y^n|U^n, X^n, Z^n, T=0) + \Pr(T=1) \cdot H(Y^n|U^n, X^n, Z^n, T=1) + h(\delta_\epsilon)$$
$$\leq n\delta_\epsilon \log |\mathcal{Y}| + H(Y^n|U^n, X^n, Z^n, T=1) + h(\delta_\epsilon)$$
$$= \sum_{(u^n, x^n, z^n) \in \mathcal{T}_\epsilon} p(u^n, x^n, z^n | T=1) H(Y^n | U^n = u^n, X^n = x^n, Z^n = z^n, T=1) + n\delta_\epsilon \log |\mathcal{Y}| + h(\delta_\epsilon)$$
$$\leq \sum_{(u^n, x^n, z^n) \in \mathcal{T}_\epsilon} p(u^n, x^n, z^n | T=1) \log |T_\epsilon^{(n)}(Y|u^n, x^n, z^n)| + n\delta_\epsilon \log |\mathcal{Y}| + h(\delta_\epsilon)$$
$$\leq n[H(Y|U, X, Z) + \delta'_\epsilon],$$

where the last inequality follows from properties of typical sequences (see [35, Ch.2]).

## APPENDIX D

### PROOF OF CONVERSE FOR ONE-SIDED HELPER

*Proof of Converse:* For any achievable tuple $(R_1, R_2, \triangle)$, by standard properties of the entropy function, it follows that

$$n(R_1 + \delta_n) \geq \log |\mathcal{W}_1^{(n)}|$$
$$\geq H(W_1) \geq H(W_1|W_2)$$
$$= H(X^n, W_1|W_2) - H(X^n|W_1, W_2)$$
$$\overset{(a)}{\geq} H(X^n, W_1|W_2) - nD$$
$$\geq \sum_{i=1}^n H(X_i|W_2, X^{i-1}) - nD$$
$$\overset{(b)}{=} \sum_{i=1}^n H(X_i|U_i) - nD$$

where $(a)$ follows from the fact that under logarithmic loss distortion $D \geq E[d(X^n, g(W_1, W_2))] \geq \frac{1}{n} H(X^n|W_1, W_2)$ (Lemma 2) and $(b)$ follows by defining $U_i \triangleq (W_2, X^{i-1})$.

Next,

$$n(R_2 + \delta_n) \geq H(W_2)$$

$$\geq I(W_2; X^n, Y^n)$$

$$= \sum_{i=1}^{n} H(X_i, Y_i) - H(X_i, Y_i | W_2, X^{i-1}, Y^{i-1})$$

$$\geq \sum_{i=1}^{n} H(X_i, Y_i) - H(X_i, Y_i | U_i)$$

$$\geq \sum_{i=1}^{n} I(Y_i; U_i).$$

Lastly, the leakage rate

$$n(\triangle + \delta_n) \geq I(X^n; W_2, Z^n)$$

$$= \sum_{i=1}^{n} H(X_i) - H(X_i | W_2, X^{i-1}, Z^n)$$

$$\geq \sum_{i=1}^{n} H(X_i) - H(X_i | U_i, Z_i).$$

We proceed by using the standard time-sharing argument. Let $Q$ be a random variable uniformly distributed over the set $\{1, 2, \ldots, n\}$ and independent of $X_i, Y_i, Z_i$, $1 \leq i \leq n$. We consider the joint distribution of new random variables $(X, Y, Z, U)$, where $X \triangleq X_Q, Y \triangleq Y_Q, Z \triangleq Z_Q$, and $U \triangleq (Q, U_Q)$. Note that we have $P_{X,Y,Z} = P_{X_Q, Y_Q, Z_Q}$ and $U - Y - (X, Z)$ forms a Markov chain due to the i.i.d. property of the source and side information sequences.

By introducing $Q$ in above expressions, it is straightforward to show that rate and leakage rate constraints above can be bounded further by

$$R_1 + \delta_n \geq H(X|U) - D$$

$$R_2 + \delta_n \geq I(Y; U)$$

$$\triangle + \delta_n \geq I(X; U, Z),$$

for some $P_{X,Y,Z} P_{U|Y}$. The proof is concluded by letting $n \to \infty$.

For the bound on the cardinality of the set $\mathcal{U}$, it can be shown by using the support lemma [32] that it suffices that $\mathcal{U}$ should have $|\mathcal{Y}| - 1$ elements to preserve $P_Y$, plus three more for $H(Y|U), H(X|U)$, and $H(X|U, Z)$.

APPENDIX E

PROOF OF THEOREM 3

With the assumption that $Y \sim \mathcal{N}(0, \sigma_Y^2)$, $X = Y + N_1, N_1 \sim \mathcal{N}(0, \sigma_{N_1}^2)$ independent of $Y$, and $Z = X + N_2, N_2 \sim \mathcal{N}(0, \sigma_{N_2}^2)$ independent of $X, Y, N_1$, we will prove that the inner bound given in Theorem 1 is tight for this case.

*Proof of Achievability*: Let us choose $U = Y + Q, Q \sim \mathcal{N}(0, \frac{\alpha}{1-\alpha}\sigma_Y^2)$ independent of $Y$, and $V = X + P, P \sim \mathcal{N}(0, \sigma_P^2)$ independent of $X$, where $\alpha \in (0,1)$ and $\sigma_P^2 = \frac{(\alpha\sigma_Y^2 + \sigma_{N_1}^2)D}{\alpha\sigma_Y^2 + \sigma_{N_1}^2 - D}$ for $D < \alpha\sigma_Y^2 + \sigma_{N_1}^2$, otherwise setting $V$ constant. Also, choose $g(U,V)$ to be an MMSE estimate of $X$ given $U$ and $V$.

With these choices of $U, V$ and $g(\cdot)$, it can be shown that

$$I(Y;U) = h(U) - h(U|Y)$$
$$= \frac{1}{2}\log(2\pi e(\sigma_Y^2 + \frac{\alpha}{1-\alpha}\sigma_Y^2)) - \frac{1}{2}\log(2\pi e(\frac{\alpha}{1-\alpha}\sigma_Y^2))$$
$$= \frac{1}{2}\log(1/\alpha),$$

and

$$I(X;V|U) = h(X|U) - h(X|U,V)$$
$$= \frac{1}{2}\log(\frac{\text{var}(X|U)}{\text{var}(X|U,V)})$$
$$= \frac{1}{2}\log(\frac{\alpha\sigma_Y^2 + \sigma_{N_1}^2}{D}),$$

for $D < \alpha\sigma_Y^2 + \sigma_{N_1}^2$, where $\text{var}(X|U) = \alpha\sigma_Y^2 + \sigma_{N_1}^2$ and $\text{var}(X|U,V) = \frac{\text{var}(X|U)\sigma_P^2}{\text{var}(X|U) + \sigma_P^2}$, and

$$I(X;U,Z) = h(X) - h(X|U,Z)$$
$$= \frac{1}{2}\log(\frac{\sigma_X^2}{\text{var}(X|U,Z)})$$
$$= \frac{1}{2}\log\Big(\frac{(\sigma_Y^2 + \sigma_{N_1}^2)(\alpha\sigma_Y^2 + \sigma_{N_1}^2 + \sigma_{N_2}^2)}{\sigma_{N_2}^2(\alpha\sigma_Y^2 + \sigma_{N_1}^2)}\Big)$$

where $\text{var}(X|U,Z) = \frac{\text{var}(X|U)\sigma_{N_2}^2}{\text{var}(X|U) + \sigma_{N_2}^2}$, and lastly,

$$E[d(X, g(U,V))] = E[(X - g(U,V))^2]$$
$$= \text{var}(X|U,V)$$
$$= D.$$

*Proof of Converse*: From the problem formulation, the joint pmf $P_{X^n, Y^n, Z^n, W_1, W_2, \hat{X}^n}$ is given by

$$P_{X^n, Y^n} P_{Z^n|X^n} P_{W_1|X^n} P_{W_2|Y^n} 1_{\{\hat{X}^n = g(W_1, W_2)\}}.$$

It follows that

$$n(R_2 + \delta_n) \geq H(W_2)$$
$$= I(W_2; Y^n)$$
$$= h(Y^n) - h(Y^n|W_2)$$
$$= n/2 \log(2\pi e \sigma_Y^2) - h(Y^n|W_2)$$
$$\stackrel{(a)}{\geq} n/2 \log(2\pi e \sigma_Y^2) - n/2 \log(2^{\frac{2}{n}h(X^n|W_2)} - 2^{\frac{2}{n}h(N_1^n|W_2)})$$
$$= n/2 \log(2\pi e \sigma_Y^2) - n/2 \log(2^{\frac{2}{n}h(X^n|W_2)} - 2\pi e \sigma_{N_1}^2),$$

where $(a)$ follows from the conditional EPI and the fact that $X^n = Y^n + N_1^n$, $Y^n$ conditionally independent of $N_1^n$ given $W_2$.

Next, consider the Markov chain $W_2 - Y^n - X^n - Z^n$, we have that

$$n/2 \log(2\pi e \sigma_X^2) = h(X^n) \geq h(X^n|W_2) \geq h(X^n|Y^n) = h(N_1^n) = n/2 \log(2\pi e \sigma_{N_1}^2).$$

Then there must exists $\alpha \in [0, 1]$ such that $h(X^n|W_2) = n/2 \log(2\pi e(\alpha \sigma_X^2 + (1-\alpha)\sigma_{N_1}^2)) = n/2 \log(2\pi e(\alpha \sigma_Y^2 + \sigma_{N_1}^2))$. Thus, we have

$$n(R_2 + \delta_n) \geq n/2 \log(2\pi e \sigma_Y^2) - n/2 \log(2\pi e \alpha \sigma_Y^2) = n/2 \log(1/\alpha).$$

Next,

$$n(R_1 + \delta_n) \geq H(W_1)$$
$$\geq I(W_1; X^n|W_2)$$
$$= h(X^n|W_2) - h(X^n|W_1, W_2)$$
$$\geq h(X^n|W_2) - \sum_{i=1}^n h(X_i|W_1, W_2)$$
$$\geq h(X^n|W_2) - \sum_{i=1}^n \frac{1}{2} \log(2\pi e \mathrm{var}(X_i|W_1, W_2))$$
$$\stackrel{(a)}{\geq} h(X^n|W_2) - \sum_{i=1}^n \frac{1}{2} \log(2\pi e E[(X_i - \hat{X}_i(W_1, W_2))^2])$$
$$\stackrel{(b)}{\geq} n/2 \log(2\pi e(\alpha \sigma_Y^2 + \sigma_{N_1}^2)) - n/2 \log(\frac{2\pi e}{n} \sum_{i=1}^n E[(X_i - \hat{X}_i(W_1, W_2))^2])$$
$$\geq n/2 \log(2\pi e(\alpha \sigma_Y^2 + \sigma_{N_1}^2)) - n/2 \log(2\pi e D)$$
$$= n/2 \log(\frac{\alpha \sigma_Y^2 + \sigma_{N_1}^2}{D})$$

where $(a)$ follows from the fact that $\mathrm{var}(X_i|W_1, W_2)$ is the MMSE over all possible estimator of $X_i$ for each $i = 1, \ldots, n$, $(b)$ follows from substituting $h(X^n|W_2) = n/2 \log(2\pi e(\alpha \sigma_X^2 + (1-\alpha)\sigma_{N_1}^2))$, and using Jensen's inequality and the fact that $\log(\cdot)$ is a concave function.

Lastly,

$$n(\triangle + \delta_n) \geq I(X^n; W_2, Z^n)$$
$$= h(X^n) - h(X^n|W_2, Z^n)$$
$$= h(X^n) - h(X^n, Z^n|W_2) + h(Z^n|W_2)$$
$$\stackrel{(a)}{=} h(X^n) - h(X^n|W_2) - h(Z^n|X^n) + h(Z^n|W_2)$$
$$\stackrel{(b)}{\geq} h(X^n) - h(X^n|W_2) - h(Z^n|X^n) + n/2 \log(2^{\frac{2}{n}h(X^n|W_2)} + 2^{\frac{2}{n}h(N_2^n|W_2)})$$
$$\stackrel{(c)}{=} n/2 \log \Big(\frac{(\sigma_Y^2 + \sigma_{N_1}^2)(\alpha\sigma_Y^2 + \sigma_{N_1}^2 + \sigma_{N_2}^2)}{\sigma_{N_2}^2(\alpha\sigma_Y^2 + \sigma_{N_1}^2)}\Big)$$

where $(a)$ follows from the Markov chain $Z^n - X^n - W_2$, $(b)$ follows from the conditional EPI and the fact that $Z^n = X^n + N_2^n$, $X^n$ conditionally independent of $N_2^n$ given $W_2$, $(c)$ follows from substituting $h(X^n|W_2)$.

## APPENDIX F
### PROOF OF CONVERSE FOR TRIANGULAR SETTING (A)

*Proof of Converse*: For any achievable tuple $(R_1, R_2, R_3, D, \triangle)$, by standard properties of the entropy function, it follows that

$$n(R_1 + R_3 + \delta_n) \geq H(W_1, W_3)$$
$$\geq I(X^n; W_1, W_3|Y^n, Z^n)$$
$$= H(X^n|Y^n, Z^n) - H(X^n|W_1, W_3, Y^n, Z^n)$$
$$\stackrel{(a)}{=} H(X^n|Y^n) - H(X^n|W_1, W_2, W_3, Y^n, Z^n)$$
$$\geq H(X^n|Y^n) - H(X^n|W_2, W_3, Y^n)$$
$$\stackrel{(b)}{\geq} H(X^n|Y^n) - nD$$
$$= \sum_{i=1}^n H(X_i|Y_i) - nD$$

where $(a)$ follows from the Markov chains $W_2 - (W_1, Z^n) - (W_3, X^n, Y^n)$ and $X^n - Y^n - Z^n$, $(b)$ follows from the fact that $D \geq E[d(X^n, g(W_2, W_3, Y^n))] \geq \frac{1}{n}H(X^n|W_2, W_3, Y^n)$ under the logarithmic loss distortion (Lemma 2).

Next

$$n(R_2 + R_3 + \delta_n) \geq H(W_2, W_3)$$
$$\geq I(W_2, W_3; X^n|Y^n)$$
$$= H(X^n|Y^n) - H(X^n|W_2, W_3, Y^n)$$
$$\geq \sum_{i=1}^n H(X_i|Y_i) - nD.$$

and the leakage rate

$$
\begin{aligned}
n(\triangle + \delta_n) &\geq I(X^n; W_1, Z^n) \\
&= I(X^n; Z^n) + I(X^n; W_1|Z^n) \\
&\stackrel{(a)}{=} I(X^n; Z^n) + H(W_1|Z^n) - H(W_1|X^n, Y^n, Z^n) \\
&\geq I(X^n; Z^n) + H(W_1|Y^n, Z^n) - H(W_1|X^n, Y^n, Z^n) \\
&\geq I(X^n; Z^n) + I(X^n; W_1|Y^n, Z^n) \\
&= I(X^n; Z^n) + I(X^n; W_1, W_3|Y^n, Z^n) - I(X^n; W_3|W_1, Y^n, Z^n) \\
&\geq I(X^n; Z^n) + I(X^n; W_1, W_3|Y^n, Z^n) - H(W_3) \\
&\stackrel{(b)}{\geq} \sum_{i=1}^{n} I(X_i; Z_i) + H(X_i|Y_i) - D - n(R_3 + \delta_n)
\end{aligned}
$$

where $(a)$ follows from the Markov chain $W_1 - X^n - (Y^n, Z^n)$ and $(b)$ follows from the steps used to bound $R_1 + R_3$, and lastly

$$
\begin{aligned}
n(\triangle + \delta_n) &\geq I(X^n; W_1, Z^n) \\
&\geq I(X^n; Z^n) \\
&= \sum_{i=1}^{n} I(X_i; Z_i)
\end{aligned}
$$

We end the proof by following the standard time-sharing argument and letting $n \to \infty$.

## APPENDIX G
## PROOF OF THEOREM 6

Since $X - Y - Z$ forms a Markov chain, we let the helper simply forward the index. Also, in the Gaussian setting with quadratic distortion, it is known that the side information at the encoder does not improve the rate distortion region, we neglect this side information in encoding. It is straightforward to show that a set of all tuples $(R_1, R_2, R_3, D, \triangle)$ satisfying the conditions below is the achievable region,

$$
\begin{aligned}
R_1 &\geq [I(X; U|Y) - R_3]^+, \\
R_2 &\geq [I(X; U|Y) - R_3]^+, \\
D &\geq E[d(X, g(U, Y))], \\
\triangle &\geq I(X; Z) + [I(X; U|Y) - R_3]^+
\end{aligned}
$$

for some $P_{X,Y} P_{Z|Y} P_{U|X}$ and $g(\cdot)$.

With the assumption that $Y = X + N_1, N_1 \sim \mathcal{N}(0, \sigma_{N_1}^2)$ independent of $X$, and $Z = Y + N_2, N_2 \sim \mathcal{N}(0, \sigma_{N_2}^2)$ independent of $X, Y, N_1$, we will prove that the achievable region above is tight for this case.

34*Proof of Achievability*: Let us choose $U = X + Q, Q \sim \mathcal{N}(0, \sigma_Q^2)$ independent of $X$, where $\sigma_Q^2 = \frac{\sigma^2 D}{\sigma^2 - D}, \sigma^2 = \frac{\sigma_X^2 \sigma_{N_1}^2}{\sigma_X^2 + \sigma_{N_1}^2}$. Also, choose $g(U, Y)$ to be an MMSE estimate of $X$ given $U$ and $Y$.

With these choices of $U$ and $g(\cdot)$, it can be shown that

$$I(X;U|Y) = h(U|Y) - h(U|X,Y)$$
$$= h(U|Y) - h(U|X)$$
$$= \frac{1}{2}\log(\frac{\sigma_Q^2 + \sigma^2}{\sigma_Q^2})$$
$$= \frac{1}{2}\log(\frac{\sigma^2}{D}),$$

and

$$I(X;Z) = h(Z) - h(Z|X)$$
$$= \frac{1}{2}\log(\frac{\sigma_X^2 + \sigma_{N_1}^2 + \sigma_{N_2}^2}{\sigma_{N_1}^2 + \sigma_{N_2}^2}),$$

and lastly

$$E[d(X, g(U,Y))] = E[(X - g(U,Y))^2]$$
$$= \text{var}(X|U,Y)$$
$$= \frac{\sigma^2 \sigma_Q^2}{\sigma^2 + \sigma_Q^2} = D$$

where $\text{var}(X|U,Y) = \frac{\text{var}(X|Y)\sigma_Q^2}{\text{var}(X|Y) + \sigma_Q^2}$.

*Proof of Converse*: From the problem formulation the joint pmf $P_{X^n, Y^n, Z^n, W_1, W_2, W_3, \hat{X}^n}$ is given by

$$P_{X^n, Y^n} P_{Z^n|Y^n} P_{W_1|X^n} P_{W_3|X^n} P_{W_2|W_1, Z^n} 1_{\{\hat{X}^n = g(W_2, W_3, Y^n)\}}.$$

35It follows that

$$n(R_1 + R_3 + \delta_n) \geq H(W_1, W_3)$$
$$\geq I(W_1, W_3; X^n | Y^n, Z^n)$$
$$\stackrel{(a)}{=} h(X^n|Y^n) - h(X^n|W_1, W_3, W_2, Y^n, Z^n)$$
$$\geq h(X^n|Y^n) - \sum_{i=1}^{n} h(X_i|W_2, W_3, Y^n)$$
$$\geq h(X^n|Y^n) - \sum_{i=1}^{n} \frac{1}{2}\log(2\pi e \text{var}(X_i|W_2, W_3, Y^n))$$
$$\stackrel{(b)}{\geq} h(X^n|Y^n) - \sum_{i=1}^{n} \frac{1}{2}\log(2\pi e E[(X_i - \hat{X}_i(W_2, W_3, Y^n))^2])$$
$$\stackrel{(c)}{\geq} n/2\log(2\pi e \sigma^2) - n/2\log(\frac{2\pi e}{n}\sum_{i=1}^{n} E[(X_i - \hat{X}_i(W_2, W_3, Y^n))^2])$$
$$\geq n/2\log(2\pi e \sigma^2) - n/2\log(2\pi e D)$$
$$= n/2\log(\frac{\sigma^2}{D})$$

where $(a)$ follows from the Markov chain $X^n - Y^n - Z^n$ and the Markov chain $W_2 - (W_1, Z^n) - (W_3, X^n, Y^n)$, $(b)$ follows from the fact that $\text{var}(X_i|W_2, W_3, Y^n)$ is the MMSE over all possible estimator of $X_i$ for each $i = 1, \ldots, n$, $(c)$ follows from Jensen's inequality and the fact that $\log(\cdot)$ is a concave function.

$$n(R_2 + R_3 + \delta_n) \geq H(W_2, W_3)$$
$$\geq I(W_2, W_3; X^n|Y^n)$$
$$\geq h(X^n|Y^n) - \sum_{i=1}^{n} h(X_i|W_2, W_3, Y^n)$$
$$\stackrel{(a)}{\geq} n/2\log(\frac{\sigma^2}{D})$$

where $(a)$ follows from steps used to prove the constraint on $R_1 + R_3$.

Lastly,

$$n(\triangle + \delta_n) \geq I(X^n; W_1, Z^n)$$
$$= I(X^n; Z^n) + I(X^n; W_1|Z^n)$$
$$\stackrel{(a)}{=} I(X^n; Z^n) + I(X^n, Y^n; W_1|Z^n)$$
$$\geq I(X^n; Z^n) + I(X^n; W_1|Y^n, Z^n)$$
$$= I(X^n; Z^n) + I(X^n; W_1, W_3|Y^n, Z^n) - I(X^n; W_3|W_1, Y^n, Z^n)$$
$$\geq I(X^n; Z^n) + I(X^n; W_1, W_3|Y^n, Z^n) - H(W_3)$$
$$\stackrel{(b)}{\geq} n/2 \log(\frac{\sigma_X^2 + \sigma_{N_1}^2 + \sigma_{N_2}^2}{\sigma_{N_1}^2 + \sigma_{N_2}^2}) + n/2 \log(\frac{\sigma^2}{D}) - n(R_3 + \delta_n)$$

where $(a)$ follows from the Markov chain $W_1 - (X^n, Z^n) - Y^n$, $(b)$ follows from steps used to prove the constraint on $R_1 + R_3$.

The constraint $\triangle + \delta_n \geq 1/2 \log(1 + \frac{\sigma_X^2}{\sigma_{N_1}^2 + \sigma_{N_2}^2})$ follows straightforwardly from $n(\triangle + \delta_n) \geq I(X^n; Z^n)$.

## APPENDIX H
### PROOF OF ACHIEVABILITY FOR TRIANGULAR SETTING (B)

The proof follows standard random coding arguments where we show the existence of the code that satisfies the rate, distortion, and leakage rate constraints. The outline of the proof is given in the following.

Codebook generation: Fix $P_{U|X}$, and the function $\tilde{g}: \mathcal{U} \times \mathcal{Y} \to \hat{\mathcal{X}}$.

- Randomly generating $2^{n(I(X;U)+\epsilon)}$ codewords $u^n(w) \sim \prod_{i=1}^n P_U(u_i(w)), w \in [1 : 2^{n(I(X;U)+\epsilon)}]$.
- Then distributed them uniformly at random into $2^{n(I(X;U|Y)+2\epsilon)}$ bins $b_U(w_u), w_u \in [1 : 2^{n(I(X;U|Y)+2\epsilon)}]$.
- We split the bin indices $w_u$ into $w_{u,1} \in [1 : 2^{n(I(X,U|Y)-R_3+\epsilon)}]$ and $w_{u,3} \in [1 : 2^{n(R_3+\epsilon)}]$.
- For secret key generation codebook, we randomly and uniformly partition the set of sequences $\mathcal{Y}^n$ into $2^{nR_k}$ bins $b_K(k), k \in [1 : 2^{nR_k}]$, where $R_k = \min\{H(Y|X,Z), H(X|Y) - D - R_3\} - 2\delta_\epsilon$.

The codebooks are revealed to the encoder, the helper, the decoder, and the eavesdropper. We consider the following two cases.

I) If $H(X|Y) < D$, we do not need to send anything over the rate-limited links. Since the decoder knows $y^n$, it can generate $\hat{x}^n$ based on $y^n$. Since $(x^n, y^n)$ are jointly typical, it can be shown that this is sufficient to satisfy the distortion, i.e., under the logarithmic loss distortion, we have $E[d(X, g(Y))] = H(X|Y)$.

II) If $H(X|Y) - D > 0$: We further split the bin indices $w_{u,1}$ into $w_{u,1k} \in [1 : 2^{nR_k}]$ and $w_{u,1l} \in [1 : 2^{n(I(X,U|Y)-R_3-R_k+\epsilon)}]$. Note that this is possible if $R_k \leq I(X,U|Y) - R_3 + \epsilon$. Note also that $w_{u,1}$ can be deduced from $(w_{u,1k}, w_{u,1l})$.

Encoding at the encoder:

- Given sequences $(x^n, y^n)$, the encoder looks for $u^n$ that is jointly typical with $x^n$. If there is more than one, it selects one of them uniformly at random. If there is no such $u^n$, it selects one out of $2^{n(I(X;U)+\epsilon)}$ uniformly at random. With high probability, there exists such $u^n$ since there are $2^{n(I(X;U)+\epsilon)}$ codewords $u^n$ generated.

- Then the encoder wishes to transmit the corresponding bin index $w_u$ to the helper and decoder in a secure way by incorporating the secret key, e.g., using "one-time pad" based on the key. To generate a secret key, the encoder looks for an index $k$ for which $y^n \in b_K(k)$. Then the encoder transmits $w_{u,1k} \oplus k$ and $w_{u,1l}$ to the helper over the cascade link, where $w_{u,1k} \oplus k$ denotes the modulo operation, $(w_{u,1k} + k) \mathrm{mod} 2^{nR_k}$, and also transmit $w_{u,3}$ to the decoder over the private (triangular) link. The helper simply forwards the indices $w_{u,1k} \oplus k$ and $w_{u,1l}$ to the decoder.

Decoding at the decoder: Upon receiving $w_{u,1k} \oplus k$, $w_{u,1l}$, and $w_{u,3}$, the decoder uses its side information $y^n$ to generate its own key and decrypt the index $w_{u,1k}$, and thus the bin index $w_u$. Then it looks for a unique $u^n$ that is jointly typical with $y^n$. With high probability, it will find the unique and correct one since there are $2^{n(I(Y;U)-\epsilon)}$ codewords in each bin $b_U(w_u)$. The decoder puts out $\hat{x}^n$ where $\hat{x}_i = g(u_i, y_i)$.

Analysis of distortion: Since $(x^n, y^n, u^n)$ are jointly typical, we can show that $D$ satisfying $D \geq E[d(X, g(U,Y))]$ is achievable. Also, due to the property of log-loss distortion function (Lemma 1), we have that $E[d(X, g(U,Y))] = H(X|U,Y)$. We define $U = X$ with probability $p = 1 - \frac{D}{H(X|Y)}$ and a constant otherwise. This gives us $H(X|U,Y) = (1-p)H(X|Y) = D$.

Analysis of leakage: The leakage averaged over all codebooks $\mathcal{C}_n$,

$$I(X^n; W_{u,1l}, W_{u,1k} \oplus K, Z^n | \mathcal{C}_n)$$
$$= I(X^n; Z^n) + I(X^n; W_{u,1l} | Z^n, \mathcal{C}_n) + I(X^n; W_{u,1k} \oplus K | W_{u,1l}, Z^n, \mathcal{C}_n)$$
$$\leq I(X^n; Z^n) + H(W_{u,1l} | \mathcal{C}_n) + H(W_{u,1k} \oplus K | \mathcal{C}_n) - H(W_{u,1k} \oplus K | W_{u,1l}, X^n, Z^n, \mathcal{C}_n)$$
$$\stackrel{(a)}{=} I(X^n; Z^n) + H(W_{u,1l} | \mathcal{C}_n) + H(W_{u,1k} \oplus K | \mathcal{C}_n) - H(K | X^n, Z^n, \mathcal{C}_n)$$
$$\stackrel{(b)}{\leq} n[I(X;Z) + H(X|Y) - D - R_3 - R_k + \epsilon + R_k] - H(K | X^n, Z^n, \mathcal{C}_n)$$
$$\stackrel{(c)}{=} n[I(X;Z) + H(X|Y) - D - R_3 - R_k + \epsilon + R_k] - I(K; Y^n | X^n, Z^n, \mathcal{C}_n)$$
$$\stackrel{(d)}{\leq} n[I(X;Z) + H(X|Y) - D - R_3 - R_k + \delta'_\epsilon]$$
$$\leq n[\triangle + \delta'_\epsilon]$$

if $\triangle \geq I(X;Z) + H(X|Y) - D - R_3 - R_k$, where $(a)$ follows from the fact that $W_{u,1l}, W_{u,1k}$ are functions of $X^n$ and $\mathcal{C}_n$, $(b)$ follows from the codebook generation, and $(c)$ follows from the fact that $K$ is a function of $Y^n$ and $\mathcal{C}_n$, and $(d)$ follows from bounding the term $H(Y^n | X^n, Z^n, K, \mathcal{C}_n)$ ( [35, lemma 22.3] when $\tilde{R} := 0$), given that $R_k < H(Y|X,Z) - \delta_\epsilon$ which holds due to the assumption that $R_k = \min\{H(Y|X,Z), H(X|Y) - D - R_3\} - 2\delta_\epsilon$ in the beginning.

## APPENDIX I

### PROOF OF CONVERSE FOR TRIANGULAR SETTING (B)

*Proof of Converse*: For any achievable tuple $(R_1, R_2, R_3, D, \triangle)$, the constraints on $R_1 + R_3$ and $R_2 + R_3$ follow the proof of triangular setting (A) (Appendix F). As for the leakage rate, we have

$$\begin{aligned}
n(\triangle + \delta_n) &\geq I(X^n; W_1, Z^n) \\
&= I(X^n; Z^n) + I(X^n; W_1 | Z^n) \\
&= I(X^n; Z^n) + I(X^n; W_1, Y^n | Z^n) - I(X^n; Y^n | W_1, Z^n) \\
&= I(X^n; Z^n) + I(X^n; Y^n | Z^n) + I(X^n; W_1 | Y^n, Z^n) - I(X^n; Y^n | W_1, Z^n) \\
&\geq I(X^n; Z^n) - H(Y^n | X^n, Z^n) + I(X^n; W_1 | Y^n, Z^n) \\
&= I(X^n; Z^n) - H(Y^n | X^n, Z^n) + I(X^n; W_1, W_3 | Y^n, Z^n) - I(X^n; W_3 | W_1, Y^n, Z^n) \\
&\geq I(X^n; Z^n) - H(Y^n | X^n, Z^n) + I(X^n; W_1, W_3 | Y^n, Z^n) - H(W_3) \\
&\overset{(a)}{\geq} \sum_{i=1}^{n} I(X_i; Z_i) - H(Y_i | X_i, Z_i) + H(X_i | Y_i) - D - n(R_3 + \delta_n)
\end{aligned}$$

where $(a)$ follows from the steps used to bound $R_1 + R_3$.

Also,

$$\begin{aligned}
n(\triangle + \delta_n) &\geq I(X^n; W_1, Z^n) \\
&= I(X^n; Z^n) \\
&= \sum_{i=1}^{n} I(X_i; Z_i).
\end{aligned}$$

We end the proof by following the standard time-sharing argument and letting $n \to \infty$.

## APPENDIX J

### PROOF OF THEOREM 9

*Proof of Achievability*: The rate and distortion constraints are the same as in the Gaussian triangular example in Setting (A). The proof follows Wyner's partitioning approach for the Gaussian Wyner-Ziv problem [34]. The leakage rate constraint however requires some new analysis. We will first show that the leakage rate $\triangle$ satisfying $\triangle > I(X, Z)$ is achievable.

We note that the side information $Y^n$ is distributed according to Gaussian distribution on $\mathbb{R}^n$. Let $\mathcal{X}_p, \mathcal{Y}_p, \mathcal{Z}_p$ be discrete sets corresponding to the partitioned version of $\mathcal{X}, \mathcal{Y}, \mathcal{Z}$ to mutually exclusive sets whose union is the entire set. Following the argument in [36, Ch.8], as we can partition $\mathbb{R}$ as fine as we wish, there exist finite partitions on $\mathcal{Y}, \mathcal{X}, \mathcal{Z}$ such that $H(Y_p | X_p, Z_p)$ can be made arbitrarily large. For example, there exist finite partitions such that

$$H(Y_p | X_p, Z_p) \geq I(X_p; U_p | Y_p) - R_3, \tag{7}$$

where $I(X_p; U_p|Y_p) - R_3$ is a term associated with the source description rate on the cascade link (Wyner-Ziv rate) in the discrete case.

The remaining proof steps follows similarly to those of the proof for Theorem 8. To generate a secret key, we randomly and uniformly partition the set $\mathcal{Y}_p^n$ into $2^{nR_k}$ bins $\mathcal{B}(k)$, where $k$ is the bin index, and we set $R_k = I(X_p; U_p|Y_p) - R_3 - 2\delta_\epsilon$. At the encoder and the decoder, given $y^n \in \mathcal{Y}^n$ which is mapped to $y_p^n \in \mathcal{Y}_p^n$, the secret key is chosen to be the bin index $k$ where $y_p^n \in \mathcal{B}(k)$. Note that, with this key rate, we are able to scramble *essentially* the whole source description $w_1$. For example, we may consider splitting the source description (Wyner-Ziv index) $w_1 \in [1 : 2^{n(I(X_p;U_p|Y_p)-R_3+\delta_\epsilon)}]$ into two parts, $w_{1,l} \in [1 : 2^{3n\delta_\epsilon}]$, and $w_{1,k} \in [1 : 2^{nR_k}]$, and transmit $w_{1,l}$ and $w_{1,k} \oplus k$ to the helper, where $w_{1,k} \oplus k$ denotes the modulo operation $(w_{1,k} + k) \mod 2^{nR_k}$.

To analyze the leakage rate averaged over all codebooks $\frac{1}{n}I(X^n; W_{1,l}, W_{1,k} \oplus K, Z^n|\mathcal{C}_n)$, we first argue that, for any $\epsilon' > 0$, there exist finite partitions of $\mathcal{X}, \mathcal{Y}$, and $\mathcal{Z}$ such that $\frac{1}{n}I(X_p^n; W_{1,l}, W_{1,k} \oplus K, Z_p^n|\mathcal{C}_n) \geq \frac{1}{n}I(X^n; W_{1,l}, W_{1,k} \oplus K, Z^n|\mathcal{C}_n) - \epsilon'$. The analysis can then be done using the similar discrete proof as in the achievability proof of Theorem 8, i.e.,

$$I(X^n; W_{1,l}, W_{1,k} \oplus K, Z^n|\mathcal{C}_n)$$
$$\leq I(X_p^n; W_{1,l}, W_{1,k} \oplus K, Z_p^n|\mathcal{C}_n) + n\epsilon'$$
$$= I(X_p^n; Z_p^n) + I(X_p^n; W_{1,l}, W_{1,k} \oplus K|Z_p^n, \mathcal{C}_n) + n\epsilon'$$
$$\leq I(X_p^n; Z_p^n) + H(W_{1,l}, W_{1,k} \oplus K|\mathcal{C}_n) - H(W_{1,l}, W_{1,k} \oplus K|X_p^n, Z_p^n, \mathcal{C}_n) + n\epsilon'$$
$$\stackrel{(a)}{=} I(X_p^n; Z_p^n) + H(W_{1,l}, W_{1,k} \oplus K|\mathcal{C}_n) - H(K|X_p^n, Z_p^n, \mathcal{C}_n) + n\epsilon'$$
$$\stackrel{(b)}{\leq} n[I(X_p; Z_p) + R_k + 3\delta_\epsilon + \epsilon'] - H(K|X_p^n, Z_p^n, \mathcal{C}_n)$$
$$\stackrel{(c)}{=} n[I(X_p; Z_p) + R_k + 3\delta_\epsilon + \epsilon'] - I(K; Y_p^n|X_p^n, Z_p^n, \mathcal{C}_n)$$
$$\stackrel{(d)}{\leq} n[I(X_p; Z_p) + \delta'_\epsilon]$$
$$\stackrel{(e)}{\leq} n[I(X; Z) + \delta'_\epsilon],$$

where $(a)$ follows from the fact that $(W_{1,l}, W_{1,k})$ is a function of $X_p^n$ and $\mathcal{C}_n$, $(b)$ follows from the codebook generation, $(c)$ follows from the fact that $K$ is a function of $Y_p^n$ and $\mathcal{C}_n$, $(d)$ follows from bounding the term $H(Y_p^n|X_p^n, Z_p^n, K, \mathcal{C}_n) \leq n(H(Y_p|X_p, Z_p) - R_k + \delta_\epsilon)$ ( [35, lemma 22.3] when $\tilde{R} := 0$), given that $R_k < H(Y_p|X_p, Z_p) - \delta_\epsilon$ which holds due to the assumption that $R_k = I(X_p; U|Y_p) - R_3 - 2\delta_\epsilon$ and (7) in the beginning, and $(e)$ follows from the Markov chain $X_p - X - Z - Z_p$. With the same choice of $U_p$ as $U$ in Theorem 6, we have proved the achievability part.

*Proof of Converse*: The converse part also follows similarly that of Theorem 6, where the constraint $\triangle + \delta_n \geq 1/2 \log(1 + \frac{\sigma_X^2}{\sigma_{N_1}^2 + \sigma_{N_2}^2})$ follows straightforwardly from $n(\triangle + \delta_n) \geq I(X^n; Z^n)$.

# APPENDIX K
## PROOF OF CONVERSE FOR TRIANGULAR SETTING (C)

*Proof of Converse*: We define $U_i \triangleq (W_2, W_3, X^{i-1}, Y^{n\setminus i})$ which satisfies $U_i - X_i - Y_i$ for all $i = 1, \ldots, n$. For any achievable tuple $(R_1, R_2, R_3, D, \triangle)$, by standard properties of the entropy function, it follows that

$$n(R_1 + R_3 + \delta_n) \geq H(W_1, W_3)$$
$$\geq I(X^n, W_1, W_3 | Y^n)$$
$$= H(X^n | Y^n) - H(X^n | W_1, W_3, Y^n)$$
$$\stackrel{(a)}{=} H(X^n | Y^n) - H(X^n | W_1, W_2, W_3, Y^n)$$
$$\geq H(X^n | Y^n) - H(X^n | W_2, W_3, Y^n)$$
$$= \sum_{i=1}^{n} H(X_i | Y_i) - H(X_i | W_2, W_3, X^{i-1}, Y^n)$$
$$\stackrel{(b)}{=} \sum_{i=1}^{n} H(X_i | Y_i) - H(X_i | U_i, Y_i)$$
$$= \sum_{i=1}^{n} I(X_i; U_i | Y_i)$$

where $(a)$ follows from the Markov chain $W_2 - W_1 - (W_3, X^n, Y^n)$ and $(b)$ follows from the definition of $U_i$.

Next

$$n(R_2 + R_3 + \delta_n) \geq H(W_2, W_3)$$
$$\geq I(W_2, W_3; X^n | Y^n)$$
$$= H(X^n | Y^n) - H(X^n | W_2, W_3, Y^n)$$
$$\geq \sum_{i=1}^{n} H(X_i | Y_i) - H(X_i | U_i, Y_i)$$
$$= \sum_{i=1}^{n} I(X_i; U_i | Y_i).$$

For the bound on distortion, we have

$$D + \delta_n \geq \frac{1}{n} \sum_{i=1}^{n} E[d(X_i, g_i^{(n)}(W_2, W_3, Y^n))]$$
$$\geq \frac{1}{n} \sum_{i=1}^{n} E[d(X_i, g_i(U_i, Y_i))],$$

and lastly, the leakage rate

$$\begin{aligned}
n(\triangle + \delta_n) &\geq I(X^n; W_1) \\
&\stackrel{(a)}{=} I(X^n, Y^n; W_1) \\
&= I(X^n, Y^n; W_1, W_3) - I(X^n, Y^n; W_3 | W_1) \\
&\geq I(X^n; W_1, W_3 | Y^n) - H(W_3) \\
&\stackrel{(b)}{\geq} \sum_{i=1}^{n} I(X_i; U_i | Y_i) - R_3 - \delta_n,
\end{aligned}$$

where $(a)$ follows from the Markov chain $W_1 - X^n - Y^n$ and $(b)$ follows from the steps used to bound $R_1 + R_3$. We end the proof by following the standard time-sharing argument and letting $n \to \infty$.

For the bound on the cardinality of the set $\mathcal{U}$, it can be shown by using the support lemma [32] that it suffices that $\mathcal{U}$ should have $|\mathcal{X}| - 1$ elements to preserve $P_X$, plus two more for $H(X|U, Y)$ and the distortion constraint.

## APPENDIX L
## PROOF OF CONVERSE FOR TRIANGULAR SETTING (D)

*Proof of Converse*: Define $U_i \triangleq (W_2, X^{i-1}, Z^{i-1}, Y^{n\setminus i})$ and $V_i \triangleq W_3$ which satisfies $(U_i, V_i, X_i) - Z_i - Y_i$ for all $i = 1, \ldots, n$. For any achievable tuple $(R_1, R_2, R_3, D, \triangle)$, by standard properties of the entropy function, it follows that

$$\begin{aligned}
n(R_1 + \delta_n) &\geq H(W_1) \\
&\geq I(X^n, W_1 | Y^n, Z^n) \\
&= H(X^n | Y^n, Z^n) - H(X^n | W_1, Y^n, Z^n) \\
&\stackrel{(a)}{=} H(X^n | Z^n) - H(X^n | W_1, W_2, Y^n, Z^n) \\
&\geq \sum_{i=1}^{n} H(X_i | Z_i) - H(X_i | W_2, X^{i-1}, Z^{i-1}, Y^{n\setminus i}, Z_i) \\
&\stackrel{(b)}{=} \sum_{i=1}^{n} H(X_i | Z_i) - H(X_i | U_i, Z_i) \\
&= \sum_{i=1}^{n} I(X_i, U_i | Z_i)
\end{aligned}$$

where $(a)$ follows from the Markov chains $W_2 - (W_1, Z^n) - (X^n, Y^n)$ and $X^n - Z^n - Y^n$, $(b)$ follows from the definition of $U_i$.

Next,

$$n(R_2 + \delta_n) \geq H(W_2) \geq I(W_2; X^n, Z^n|Y^n)$$
$$= H(X^n, Z^n|Y^n) - H(X^n, Z^n|W_2, Y^n)$$
$$= \sum_{i=1}^{n} H(X_i, Z_i|Y_i) - H(X_i, Z_i|W_2, X^{i-1}, Z^{i-1}, Y^n)$$
$$= \sum_{i=1}^{n} H(X_i, Z_i|Y_i) - H(X_i, Z_i|U_i, Y_i)$$
$$= \sum_{i=1}^{n} I(X_i, Z_i; U_i|Y_i),$$

and

$$n(R_3 + \delta_n) \geq H(W_3) \geq I(W_3; X^n, Z^n|W_2, Y^n)$$
$$= H(X^n, Z^n|W_2, Y^n) - H(X^n, Z^n|W_2, W_3, Y^n)$$
$$= \sum_{i=1}^{n} H(X_i, Z_i|W_2, X^{i-1}, Z^{i-1}, Y^n) - H(X_i, Z_i|W_2, W_3, X^{i-1}, Z^{i-1}, Y^n)$$
$$= \sum_{i=1}^{n} H(X_i, Z_i|U_i, Y_i) - H(X_i, Z_i|U_i, V_i, Y_i)$$
$$= \sum_{i=1}^{n} I(X_i, Z_i; V_i|U_i, Y_i).$$

For the bound on distortion, we have

$$D + \delta_n \geq \frac{1}{n} \sum_{i=1}^{n} E[d(X_i, g_i^{(n)}(W_2, W_3, Y^n))]$$
$$\geq \frac{1}{n} \sum_{i=1}^{n} E[d(X_i, g_i(U_i, V_i, Y_i))],$$

and lastly, the leakage rate

$$n(\triangle + \delta_n) \geq I(X^n; W_1, Z^n)$$
$$\stackrel{(a)}{=} H(X^n) - H(X^n|W_1, Z^n, Y^n)$$
$$\stackrel{(b)}{=} H(X^n) - H(X^n|W_1, W_2, Z^n, Y^n)$$
$$= \sum_{i=1}^{n} H(X_i) - H(X_i|W_1, W_2, X^{i-1}, Z^n, Y^n)$$
$$\geq \sum_{i=1}^{n} I(X_i; U_i, Z_i)$$

where $(a)$ follows from the Markov chain $(W_1, X^n) - Z^n - Y^n$, $(b)$ follows from the Markov chain $W_2 - (W_1, Z^n) - (X^n, Y^n)$. We end the proof by following the standard time-sharing argument and letting $n \to \infty$.

For the bounds on the cardinalities of the sets $\mathcal{U}$ and $\mathcal{V}$, it can be shown by using the support lemma [32] that it suffices that $\mathcal{U}$ should have $|\mathcal{X}||\mathcal{Z}| - 1$ elements to preserve $P_{X,Z}$, plus four more for $H(X|U, Z)$, $I(X, Z; U|Y)$, $I(X, Z; V|U, Y)$, and the distortion constraint. The new variable $U$ induces the new $V$, and for each $U = u$, it suffices to consider $|\mathcal{V}| \leq |\mathcal{X}||\mathcal{Z}| + 1$ so that $P_{X,Z|U=u}$, $I(X, Z; V|U = u, Y)$, and the distortion constraint are preserved. Thus, the overall cardinality bound for $\mathcal{V}$ is $|\mathcal{V}| \leq |\mathcal{U}|(|\mathcal{X}||\mathcal{Z}| + 1) \leq (|\mathcal{X}||\mathcal{Z}| + 3)(|\mathcal{X}||\mathcal{Z}| + 1)$.

## APPENDIX M
## PROOF OF THEOREM 14

*Proof of Achievability*: The achievability proof follows from a standard random coding argument.

*Codebook Generation*: Fix $P_{U|X,Z}$ and $g : \mathcal{U} \times \mathcal{Y} \to \hat{\mathcal{X}}$. Let $\mathcal{W}_1^{(n)} = [1 : 2^{nR_1}]$, $\mathcal{W}_2^{(n)} = [1 : 2^{nR_2}]$, and $\mathcal{W}'^{(n)} = [1 : 2^{nR'}]$. The codewords $u^n(w_1, w_2, w')$ are generated i.i.d. each according to $\prod_{i=1}^n P_U(u_i)$, for $(w_1, w_2, w') \in \mathcal{W}_1^{(n)} \times \mathcal{W}_2^{(n)} \times \mathcal{W}'^{(n)}$. The codebook is then revealed to the encoder, helper, and decoder.

*Encoder*: Given a source sequence $x^n$, and side information $z^n$ the encoder first looks for $u^n(w_1, w_2, w')$ that is jointly typical with $(x^n, z^n)$. If there exists such a codeword, the encoder transmits the smallest $w_1$ to the helper and the decoder. If not successful, the encoder transmits $w_1 = 1$. By the covering lemma [35, Ch.3], the encoder is successful if $R_1 + R_2 + R' > I(X, Z; U) + \delta_\epsilon$.

*Helper*: Given an index $w_1$, and side information $z^n$ the helper looks for a unique $(w_2, w')$ such that $u^n(w_1, w_2, w')$ is jointly typical with $z^n$. If successful, the helper transmits the corresponding $w_2$ to the decoder. If not successful, the helper transmits $w_2 = 1$. By the packing lemma [35, Ch.3], the helper is successful if $R_2 + R' < I(Z; U) - \delta_\epsilon$.

*Decoder*: Given the indices $w_1$ and $w_2$, and the side information $y^n$ the decoder looks for a unique $u^n(w_1, w_2, w')$ such that it is jointly typical with $y^n$. If successful, the decoder reconstructs as $\hat{x}^n$ where $\hat{x}_i = g(u_i(w_1, w_2, w'), y_i)$. Otherwise, the decoder puts out $\hat{x}^n$ where $\hat{x}_i = g(u_i(w_1, w_2, 1), y_i)$. By the packing lemma, the decoder is successful if $R' < I(Y; U) - \delta_\epsilon$.

By combining the bounds on the code rates above, we obtain $R_1 > I(X; U|Z) + 2\delta_\epsilon$, and $R_1 + R_2 > I(X, Z; U|Y) + 2\delta_\epsilon$. Analysis of the distortion constraint follows standard arguments using the fact that $(X^n, U^n, Y^n)$ are jointly typical. The leakage analysis follows similarly as in the proof of Theorem 13. This concludes the achievability proof.

*Proof of Converse*: Define $U_i \triangleq (W_2, X^{i-1}, Z^{i-1}, Y^{n\setminus i})$ which satisfies $(U_i, X_i) - Z_i - Y_i$ for all $i = 1, \ldots, n$. The proof of constraints on $R_1, D$ and $\triangle$ follow similarly as in that of Theorem 13. As for the sum rate $R_1 + R_2$, it follows that

$$n(R_1 + R_2 + \delta_n) \geq H(W_1, W_2) \geq I(W_1, W_2; X^n, Z^n|Y^n)$$
$$= H(X^n, Z^n|Y^n) - H(X^n, Z^n|W_1, W_2, Y^n)$$
$$\geq \sum_{i=1}^n H(X_i, Z_i|Y_i) - H(X_i, Z_i|W_2, X^{i-1}, Z^{i-1}, Y^n)$$
$$= \sum_{i=1}^n H(X_i, Z_i|Y_i) - H(X_i, Z_i|U_i, Y_i)$$
$$= \sum_{i=1}^n I(X_i, Z_i; U_i|Y_i),$$

For the bounds on the cardinalities of the sets $\mathcal{U}$, it can be shown by using the support lemma [32] that it suffices that $\mathcal{U}$ should have $|\mathcal{X}||\mathcal{Z}| - 1$ elements to preserve $P_{X,Z}$, plus three more for $H(X|U,Z)$, $I(X,Z;U|Y)$, and the distortion constraint.